\documentclass[twocolumn]{aastex631}

\begin{document}

\title{The Galaxy Activity, Torus, and outflow Survey (GATOS) XIII: Coupling Driven $\rm H_2$ Excitation in Seyferts}

\author[0009-0007-6992-2555]{Daniel E. Delaney}
\affiliation{Department of Physics, University of Alaska Fairbanks, AK 99775, USA}
\affiliation{Department of Physics and Astronomy, University of Alaska Anchorage, AK 99508, USA}

\author[0000-0002-4457-5733]{Erin K. S. Hicks}
\affiliation{Department of Physics and Astronomy, University of Alaska Anchorage, AK 99508, USA}
\affiliation{Department of Physics, University of Alaska Fairbanks, AK 99775, USA}
\affiliation{Department of Physics and Astronomy, The University of Texas at
San Antonio, 1 UTSA Circle, San Antonio, TX 78249, USA}

\author[0000-0003-4937-9077]{Lulu Zhang}
\affiliation{Department of Physics and Astronomy, The University of Texas at
San Antonio, 1 UTSA Circle, San Antonio, TX 78249, USA}

\author[0000-0003-4949-7217]{Ric Davies}
\affiliation{Max Planck Institute for Extraterrestrial Physics (MPE), Giessenbachstr.1, 85748 Garching, Germany}

\author[0000-0001-7827-5758]{Chris Packham}
\affiliation{Department of Physics and Astronomy, The University of Texas at
San Antonio, 1 UTSA Circle, San Antonio, TX 78249, USA}

\author[0000-0003-0483-3723]{Rogemar A. Riffel}
\affiliation{Departamento de F\'isica, CCNE, Universidade Federal de Santa Maria, Av. Roraima 1000, 97105-900,  Santa Maria, RS, Brazil}

\author[0000-0002-4005-9619]{Miguel Pereira Santaella}
\affiliation{Instituto de Física Fundamental, CSIC, Calle Serrano 123, 28006
Madrid, Spain}

\author[0000-0001-9791-4228]{Enrica Bellocchi}
\affiliation{Departamento de Física de la Tierra y Astrofísica, Fac. de CC. Físicas, Universidad Complutense de Madrid, 28040 Madrid, Spain}
\affiliation{Instituto de Física de Partículas y del Cosmos IPARCOS, Fac. CC. Físicas, Universidad Complutense de Madrid, 28040 Madrid, Spain}

\author[0000-0003-4209-639X]{Nancy A. Levenson}
\affiliation{Space Telescope Science Institute, Baltimore, MD, US}

\author[0000-0001-9520-7765]{Steph Campbell}
\affiliation{School of Mathematics, Statistics and Physics, Newcastle University, Newcastle upon Tyne, NE1 7RU, UK}

\author[0000-0002-0001-3587]{David J. Rosario}
\affiliation{School of Mathematics, Statistics and Physics, Newcastle University, Newcastle upon Tyne, NE1 7RU, UK}

\author[0000-0002-3741-6136]{Houda Haidar}
\affiliation{School of Mathematics, Statistics and Physics, Newcastle University, Newcastle upon Tyne, NE1 7RU, UK}  

\author[0000-0001-8353-649X]{Cristina Ramos Almeida}
\affiliation{Instituto de Astrof\'isica de Canarias, Calle V\' ia L\'actea, s/n, E-38205, La Laguna, Tenerife, Spain} 
\affiliation{Departamento de Astrof\'isica, Universidad de La Laguna, E-38206, La Laguna, Tenerife, Spain}  

\author[0000-0003-3589-3294]{Anelise Audibert}
\affiliation{Instituto de Astrof\'isica de Canarias, Calle V\' ia L\'actea, s/n, E-38205, La Laguna, Tenerife, Spain} 
\affiliation{Departamento de Astrof\'isica, Universidad de La Laguna, E-38206, La Laguna, Tenerife, Spain}  

\author[0000-0001-5231-2645]{Claudio Ricci}
\affiliation{Department of Astronomy, University of Geneva, ch. d'Ecogia 16, 1290, Versoix, Switzerland} 
\affiliation{Instituto de Estudios Astrof\'isicos, Facultad de Ingenier\'ia y Ciencias, Universidad Diego Portales, Av. Ej\'ercito Libertador 441, Santiago, Chile}  

\author[0000-0002-9610-0123]{Laura Hermosa Mu{\~n}oz}
\affiliation{Centro de Astrobiolog\'ia (CAB) CSIC-INTA, Camino Bajo del Castillo s/n, 28692 Villanueva de la Ca\~nada, Madrid, Spain} 

\author[0000-0003-2658-7893]{Francoise Combes}
\affiliation{\textit{Observatoire de Paris, LUX, }\textit{Coll\`ege de France, }\textit{PSL Research University, CNRS, Sorbonne University, Paris, France}}  

\author[0000-0001-6794-2519]{Almudena Alonso-Herrero}
\affiliation{Centro de Astrobiolog\'ia (CAB) CSIC-INTA, Camino Bajo del Castillo s/n, 28692 Villanueva de la Ca\~nada, Madrid, Spain}  

\author[0000-0003-0444-6897]{Santiago Garc\'ia-Burillo}
\affiliation{Observatorio Astron\'omico Nacional (OAN-IGN)-Observatorio de Madrid, Alfonso XII, 3, 28014-Madrid, Spain}  

\author[0000-0002-5775-6285]{Federico Esposito}
\affiliation{Observatorio Astron\'omico Nacional (OAN-IGN)-Observatorio de Madrid, Alfonso XII, 3, 28014-Madrid, Spain}  

\author[0000-0002-9627-5281]{Ismael Garc\'ia-Bernete}
\affiliation{Centro de Astrobiolog\'ia (CAB), CSIC-INTA, Camino Bajo del Castillo s/n, E-28692 Villanueva de la Ca\~nada, Madrid, Spain}  

\author[0000-0002-2125-4670]{Taro Shimizu}
\affiliation{Max Planck Institute for Extraterrestrial Physics (MPE), Giessenbachstr.1, 85748 Garching, Germany}  

\author[0000-0003-1810-0889]{Martin Ward}
\affiliation{Centre for Extragalactic Astronomy, Department of Physics, Durham University, South Road, Durham, DH1 3LE, UK}  

\author[0000-0002-2356-8358]{Omaira Gonzalez Martin}
\affiliation{Instituto de Radioastronom\'ia y Astrof\'isica (IRyA), Universidad Nacional Aut\'onoma de M\'exico, Antigua Carretera a P\'atzcuaro \#8701, Colonia ExHda. San Jos\'e de la Huerta, Morelia, Michoac\'an, M\'exico C.P. 58089}  

\author[0000-0002-0690-8824]{Alvaro Labiano}
\affiliation{Telespazio UK for ESA, ESAC, Camino Bajo del Castillo s/n, 
28692 Villanueva de la Cañada, Spain.}  

\author[0009-0009-0466-9223]{Oscar Veenema}
\affiliation{Department of Physics, University of Oxford, Keble Road, Oxford, OX1 3RH, UK}  

\author[0000-0001-5357-6538]{Enrique Lopez-Rodriguez}
\affiliation{Kavli Institute for Particle Astrophysics \& Cosmology (KIPAC), Stanford University, Stanford, CA 94305, USA}  

\author[0000-0001-6854-7545]{Dimitra Rigopoulou}
\affiliation{Department of Physics, University of Oxford, Keble Road, Oxford, OX1 3RH, UK}  
\affiliation{School of Sciences, European University Cyprus, Diogenes street, Engomi, 1516 Nicosia, Cyprus}  

\author[0000-0001-5146-8330]{Marko Stalevski}
\affiliation{Astronomical Observatory, Volgina 7, 11060 Belgrade, Serbia}  
\affiliation{Sterrenkundig Observatorium, Universiteit Gent, Krijgslaan 281-S9, Gent B-9000, Belgium}  

\author[0000-0002-6353-1111]{Sebastian F. H\"onig}
\affiliation{Department of Physics \& Astronomy, University of Southampton, Highfield, Southampton SO171BJ, UK}  

\author[0000-0001-8042-9867]{Donaji Esparza-Arredondo}
\affiliation{Instituto de Radioastronom\'ia y Astrof\'isica (IRyA), Universidad Nacional Aut\'onoma de M\'exico, Antigua Carretera a P\'atzcuaro \#8701, Colonia ExHda. San Jos\'e de la Huerta, Morelia, Michoac\'an, M\'exico C.P. 58089}  

\author[0000-0001-9452-0813]{Takuma Izumi}
\affiliation{Department of Astronomy, School of Science, Graduate University for Advanced Studies (SOKENDAI), Mitaka, Tokyo 181-8588, Japan}  

\author[0000-0003-4809-6147]{Lindsay Fuller}
\affiliation{Department of Physics and Astronomy, The University of Texas at
San Antonio, 1 UTSA Circle, San Antonio, TX 78249, USA}  

\author{Daniel Rouan}
\affiliation{LESIA, Observatoire de Paris, Université PSL, CNRS, Sorbonne Université, Sorbonne Paris Citeé, 5 place Jules Janssen, F-92195 Meudon, France}  



\begin{abstract}

We utilize JWST/MIRI IFU observations from the Galaxy Activity, Torus and Outflow Survey (GATOS) to investigate the diverse range of ionized outflow rates of obscured AGN with similar bolometric luminosity and explore potential associations with AGN feedback. We explore spatial correlations between ionized emission potentially associated with fast shocks ([Fe~II]$_{5.34 \rm \mu m}$) and the excitation of $\rm H_2$. We further constrain our investigation to the inner 400\,pc (the nuclear and circumnuclear regions r~$<$~200 pc), and estimate the excitation temperature and column density of $\rm H_2$ assuming local thermodynamic equilibrium (LTE) and using the S(1) to S(8) rotational $\rm H_2$ emission lines visible to JWST/MIRI spectroscopy. We report the molecular gas temperature of the deprojected 400\,pc nuclear region to correlate with the ionized mass outflow rate. We also observe the stronger degree of spatial correlation between [Fe~II]$_{5.34um}$ emission and $\rm H_2$ gas temperature. We observe regions of enhanced [Fe~II]$_{5.34 \mu m}$ / [Ar II]$_{6.99 \mu m}$  spatially coincident with the ionization cones of objects with higher ionized outflow rate and [Fe~II]$_{5.34 \mu m}$ / [Ar II]$_{6.99 \mu m}$ in the deprojected 400\,pc nuclear region to scale positively with both ionized outflow rate and estimated molecular gas temperature. We do not observe the estimated jet cavity power within the central 400\,pc to strongly correlate with the ionized mass outflow rate or molecular gas temperature of the nuclear region.  We take the preceding observations to suggest a higher degree of interaction between AGN outflows and the circumnuclear disk.

\end{abstract}



\section{Introduction} \label{sec:intro}

Large amounts of non-stellar radiation emanating from a galactic center is the result of mass accretion around the galaxies' central supermassive black hole (SMBH). The high levels of energy associated with these active galactic nuclei (AGN) is thought to play an important role in galaxy evolution as these objects provide feedback from the growth of the SMBH to the rest of galaxy \citep{SilkRees1998, khalatyan2008agn,fabian2012observational, hickox2018}. Currently, our understanding of the anatomy of AGN is described by the AGN unified model (e.g. \citealt{antonucci1993unified, urry1995unified}) in which the central SMBH is surrounded by a luminous accretion disk of thermo-viscous plasma and at larger scales surrounded by a dusty toroidal structure shrouding the AGN. The orientation and geometry of these components relative to the circumnuclear structures such as the circumnuclear rotating disk have implications to how the ionization bicone and outflowing material provide feedback to the host galaxy. While the interplay between AGN-host galaxy is complex \citep{almeida2017nuclear, harrison2024observational}, it remains a critical component in understanding galactic evolution. In fact, AGN activity is thought to regulate both the growth of the central SMBH itself (e.g. \citealt{hopkins2005black}) as well as impact large scale galactic processes such as star formation via quenching (e.g. \citealt{fabian2012observational, cicone2014massive, harrison2018agn}) or enhancement (see \citealt{davies2007close, esquej2013nuclear, maiolino2017star, harrison2017impact, ellison2018star, bessiere2022spatially, molina2023enhanced, munoz2024biconical}).

AGN associated feedback can result from both radiative and kinetic processes and results in the redistribution of material and the provision of excitation energy \citep{veilleux2005galactic, fabian2012observational, wada2012radiation,veilleux2013fast, almeida2017nuclear, harrison2018agn, garcia2024deciphering}. In addition to AGN induced radiative heating, in Seyfert galaxies, components of the AGN outflows, such as radio jets and AGN winds (which have been shown to be relatively common \citealt{crenshaw2010geometry, crenshaw2012feedback, fischer2011hubble, fischer2018hubble, riffel2023agnifs}), can provide kinetic feedback to the host galaxy. These kinetic mechanisms have the potential to induce shocks in the AGN circumnuclear region, destroying dust, heating, and disturbing gas \citep{riffel2025impact}. As such, it is reasonable to suspect that the relative geometry of AGN components (e.g. AGN ionization-cone and circumnuclear disk coupling; \citealt{almeida2022diverse}) may regulate AGN feedback and behavior. 

In this work, we utilize Integral Field Unit (IFU) observations from the James Webb Space Telescope (JWST; \citealt{gardner2023james, rigby2023science}) Mid Infrared Instrument (MIRI; \citealt{rieke2015mid}) to investigate the excitation of $\rm H_2$ in the circumnuclear region of six Seyferts (type 1.9-2) with bolometric luminosity within an order of magnitude (in range 10$^{43.4}$-10$^{44.3}$ erg s$^{-1}$) and similar distances, but estimated ionized mass outflow rates ranging from 0.003 to 0.52 $\rm M_{\odot} \,yr^{-1}$ (derived from optical observations,\citealt{davies2020ionized}) (0.03 to 0.33 $\rm M_{\odot} \,yr^{-1}$; derived using observations in the MIR \citealt{zhang2024galaxy}). These observations of otherwise similar objects presenting a wide range of ionized mass outflow rates highlights an important puzzle of AGN behavior which must be solved to fully understand the role of the central SMBH engine to the host galaxy. Previously, it had been suggested that the range of observed ionized outflow rates, may be an artifact of differing relevant timescales of AGN luminosities ($\approx$ 10$^4$ - 10$^5$ yr) and outflow properties ($\approx$ 10$^6$ yr) \cite{zubovas2020intermittent}, or perhaps driven by geometric coupling between AGN outflows and the ambient disk promoting outflow mass loading and thereby bolstering the observed outflow rate \citep{fischer2016gemini, fischer2018hubble}. In previous work \citep{delaney2025excitation}, we explored the excitation of $\rm H_2$ in two Seyferts from the Galaxy Activity, Torus, and Outflow Survey (GATOS\footnote{More information regarding the GATOS collaboration and their ongoing research can be found at \url{https://gatos.myportfolio.com/home}.}) Cycle\,1 sample (NGC\,3081 and NGC\,5506). We identified a differing suite of excitation mechanisms and AGN interactions with the circumnuclear $\rm H_2$. For NGC\,5506, we identified warm molecular gas entrained in the AGN outflow and suspected AGN induced shock excitation to likely be an important excitation mechanism. Whereas for NGC\,3081, AGN photoionization was sufficient to describe observations of the circumnuclear $\rm H_2$. This work acts to extend this analysis of circumnuclear $\rm H_2$ along the dimension of ionized outflow rate by incorporating the full GATOS Cycle\,1 sample of six Seyferts, at a fixed physical scale. 

Utilizing the 8 rotational emission lines visible to JWST/MIRI spectroscopy we explore the spatial distribution of $\rm H_2$ temperature and its distribution in relation to [Fe~II]$_{5.34\rm \mu m}$ (associated with shocks; \citealt{koo2016infrared, pereira2024extended, alonso2025miconic}), and map the distribution of [Fe II]$_{5.34 \mu m}$ normalized by [Ar II]$_{6.99 \mu m}$ (a primarily photo-ionized emission line \citealt{Sofia1998, verma2003mid, amayo2021ionization}). In addition, we investigate the relationship between the observed ionized outflow rate and molecular gas temperature and [Fe II]$_{5.34 \mu m}$/[Ar II]$_{6.99 \mu m}$ of the nuclear region as well as the estimated radio jet cavity power. We utilize methods similar to those used in previous studies which investigated $\rm H_2$ gas population levels in a variety of objects including ultra-luminous infrared galaxies (ULIRGs), Low-ionization nuclear emission-line region (LINER), and Seyfert galaxies (see \citealt{ogle2007shocked, pereira2014warm, togi2016lighting, zhang2023interaction,davies2024gatos, herrero2024miconic, delaney2025excitation} and \citealt{hunt2025interstellar}). This work acts to build upon previous studies which have investigated winds, feedback, and molecular outflows in local Seyferts. To highlight a few, interactions between AGN ionization cones and the circumnuclear disk have been identified by \cite{davies2014fueling}, \cite{garcia2024galaxy} and \cite{esposito2024agn} and molecular outflows from variety of AGN have been identified (e.g.  \citealt{gallimore2016high, imanishi2018alma,impellizzeri2019counter,alonso2019nuclear,herrero2023agn, garcia2014molecular,garcia2016alma,garcia2019alma, garcia2021galaxy, garcia2021multiphase, garcia2024galaxy, delaney2025excitation} and \citealt{imanishi2025almasubparsecresolutiondense}).

The organization of this paper is as follows: in Section \ref{sec:Sample} we present the sample selection and describe the data collection and reduction process. In Section \ref{sec:Molecular Gas Excitation} we present the spatial distribution and molecular gas temperature, ionized, and radio emission. In Section \ref{sec:Central 400 pc} we explore the molecular gas temperature (Section \ref{subsec:Central 400pc Temperature}), [Fe II]$_{5.34 \mu m}$/[Ar II]$_{6.99 \mu m}$ (Section \ref{subsec:Central 400pc [Fe II]/[Ar II]}) and radio jet cavity power (Section \ref{subsec:Central 400pc Jet Power}) of the de-projected 400\,pc nuclear region (r $<$ 200\,pc). Finally, we provide a summary of results in Section \ref{sec:Summary}. Additional information regarding the sample and methodology and background information is presented in the Appendix \ref{Appendix A}.

\section{Sample and Observations} \label{sec:Sample}

The GATOS collaboration observed six nearby AGN in the mid-infrared (MIR) as part of the JWST Cycle\,1 program 1670 (P.I. T. Shimizu). Targets (Table \ref{Galaxy Fundamentals}) were selected from the \cite{davies2015insights} sample who identified objects in the 70-month \textit{Swift}-BAT All-sky Hard X-ray Survey \citep{baumgartner201370} as ideal candidates to investigate the behavior of AGN outflows. The GATOS Cycle\,1 sample is designed to explore differences in ionized mass outflow rates: galaxies within the sample span a narrow range of key parameters but present a wide range of observed ionized outflow rates. The sample is distance limited ($<$ 40 Mpc) and all objects have bolometric luminosities within an order of magnitude (in range 10$^{43.4}$-10$^{44.3}$ erg s$^{-1}$) but with outflow rates spanning two orders of magnitude \citep{davies2020ionized, zhang2024galaxy}. In addition, these targets are of similar AGN type (Seyfert 1.9-2) and host galaxies in the sample include 3 lenticular and 3 spiral galaxies. To this date, numerous studies have been conducted utilizing the GATOS Cycle\,1 sample set and analyzing each individual target. In Appendix \ref{subsec:Evidence_for_Winds} we provide a brief overview of each target and potential evidence for strong/weak outflow-disk coupling.

\begin{deluxetable*}{rccccccccccc}
\tabletypesize{\scriptsize}
\tablewidth{0pt} 
\tablecaption{Galaxy Fundamental Parameters.  \label{Galaxy Fundamentals}}
\tablehead{
\colhead{Galaxy} & \colhead{Host Type} & \colhead{AGN Type} & \colhead{z} & \colhead{D$_{\rm L}$} & \colhead{PA$_{\rm disk}$} & \colhead{i$_{\rm disk}$} & \colhead{PA$_{\rm cone}$} & \colhead{$\Omega_{\rm Out}$} & \colhead{$\rm \dot{M}_{out}$} & \colhead{log[L$_{AGN}$]} & \colhead{log[L$_{14-195 keV}$]} \\ 
\colhead{} & \colhead{} & \colhead{} & \colhead{} & \colhead{(Mpc)} & \colhead{(deg)} & \colhead{(deg)} & \colhead{(deg)} & \colhead{(deg)} & \colhead{($\rm M_{\odot} yr^{-1}$)} & \colhead{(erg s$^{-1}$)} & \colhead{(erg s$^{-1}$)}
} 
\colnumbers
\startdata 
\textbf{ESO 137-G034}    &   SAB0/a  &   2  &   0.00914  &   35  &   168  &   38  &   124  &   78  &   0.52; 0.33  & 43.4 & 42.77 \\
\textbf{NGC 5506}   &   Sa pec  &   1.9/1i  &   0.00608  &   27  &   265  &   80  &   22  &   80  &   0.21; 0.28  & 44.1 & 43.31 \\
\textbf{NGC 5728}   &   SAB(r)a  &   1.9  &   0.00932  &   39  &   14  &   43  &   127  &   46  &   0.09; 0.08  & 44.1  & 43.26  \\
\textbf{NGC 3081}   &   (R)SAB0/a(r)  &   2  &   0.00798  &   34  &   75  &   41  &   165  &   30  &   0.04; 0.03 & 44.1   & 43.12  \\
\textbf{NGC 7172}   &   Sa pec  &   2  &   0.00868  &   37  &   90  &   88  &   2  &   120  &   0.005; 0.03 & 44.1   & 43.48  \\
\textbf{MCG-05-23-016}    &   S0  &   2  &   0.00849  &   35  &   59  &   35  &   172  &   --  &   0.003; 0.04 &  44.3 & 43.54  \\
\enddata
\tablecomments{Column (2)-(4): Host galaxy type, and object redshift, are taken from NASA/IPAC Extragalactic Database (NED); "r" and "R" indicate the presence of an inner or outer ring respectively. Column (5): Object distances are taken from \cite{caglar2020llama}.  (6): References for the position angle of the circumnuclear rotational disk for 
ESO137-G034 taken from \cite{zhang2024galaxy};
NGC\,5506 taken from \cite{esposito2024agn};
NGC\,5728 taken from \cite{shimizu2019multiphase}; 
NGC\,3081 taken from \cite{ruschel2021agnifs}; 
NGC\,7172 taken from \cite{herrero2023agn}; 
MCG-05-23-016 taken from \cite{esparza2025molecular}.
Column (7): Inclination of the circumnuclear disk for 
ESO137-G034 taken from \cite{burtscher2021llama};
NGC\,5506 taken from \cite{esposito2024agn};
NGC\,5728 taken from \cite{shimizu2019multiphase}; 
NGC\,3081 taken from \cite{burtscher2021llama}; 
NGC\,7172 taken from \cite{herrero2023agn}; 
MCG-05-23-016 taken from \cite{esparza2025molecular}.
Column (8): Position angle of the ionization cone for
ESO137-G034 taken from \cite{zhang2024galaxy};
NGC\,5506 taken from \cite{fischer2013determining};
NGC\,5728 taken from \cite{zhang2024galaxy}; 
NGC\,3081 taken from \cite{zhang2024galaxy}; 
NGC\,7172 taken from \cite{herrero2023agn}; 
MCG-05-23-016 taken from \cite{zhang2024galaxy}.
Column (9): Ionization cone opening angle for
ESO137-G034 taken from \cite{ma2020extended}; 
NGC\,5506 taken from \cite{fischer2013determining}; 
NGC\,5728 taken from \cite{shimizu2019multiphase}; 
NGC\,3081 taken from \cite{schnorr2016feeding}; 
NGC\,7172 taken from \cite{herrero2023agn}; 
MCG-05-23-016, no information available.
Column (10): estimated ionized mass outflow rates are from \cite{davies2020ionized}; \cite{zhang2024galaxy}. Column (11): AGN bolometric luminosity from \cite{davies2020ionized}. Column (12): 14-195 keV X-ray luminosity taken from the 157-month Swift-BAT survey \cite{lien2025157}.}
\end{deluxetable*}

\subsection{Data Collection} \label{subsec:Data_Collection_Reduction}

This study utilizes MIR IFU observations from the Mid Infrared Instrument (MIRI; \citealt{rieke2015mid}) on board JWST. JWST/MIRI is capable of performing medium resolution spectroscopy (MRS; \citealt{wells2015mid}; \citealt{argyriou2023jwst}) between 4.9 to 27.9 $\rm \mu$m \citep{wells2015mid, argyriou2023jwst}. Datacubes were processed via the standard calibration pipeline (version 1.11.4) and supplemented with some additional steps masking erroneously hot and cold pixels which were not identified by the standard data reduction pipeline. The data reduction process is described in detail in \cite{pereira2022low}, \cite{garcia2024galaxy}, and \cite{garcia2022high}.  

\section{Distributions of Molecular Excitation} \label{sec:Molecular Gas Excitation}

In optically-thin gas, the observed line flux is directly proportional to the number of emitting particles. Assuming a Boltzmann's distribution, the temperature of a gas can be calculated from the observed flux ratio of two rotational hydrogen lines with corresponding energies $E_i$ and $E_j$ 
\begin{equation}
T = \frac{E_{j} - E_i}{k \cdot \ln \left( \frac{g_j A_j \lambda_i F_i }{g_i A_i \lambda_j F_j} \right) } \label{Excitation Temp Eqn}
\end{equation}
where $E_i$ and $E_j$ are the energy level corresponding with states $i$ and $j$ respectively, $k$ is Boltzmann's constant, $F_{i}$ is the observed total flux, $\lambda_i$ is the frequency of emission, $A_i$ is the probability of transition, and $g_i$ is the degeneracy value associated with the given state denoted by $i$ (or $j$ respectively). 

Using the $\rm H_2$ 0-0 S(1)/S(5) line flux ratio (line fitting described in Appendix \ref{Appendix:Line Fitting}) we map the temperature of $\rm H_2$ for each object within the sample (Figure \ref{Excitation and Ionized Emission}). The molecular gas temperature is centrally peaked near the AGN for each object, but the extended morphology varies. The three objects with the highest ionized outflow rates (ESO137-G034, NGC\,5506, and NGC\,5728) show extension aligned with the ionization cones of each object, whereas for the three objects with lower ionized outflow rates (NGC\,3081, NGC\,7172, and MCG-05-23-016) we see a more centrally concentrated temperature distribution. This suggests a potential connection between the ionization cone and extended $\rm H_2$ excitation for objects presenting increased rates of ionized outflow. 

\subsection{$\rm H_2$ Temperature Distributions} \label{subsec:Temp and [Fe II]}

\begin{figure*}[ht]
\plotone{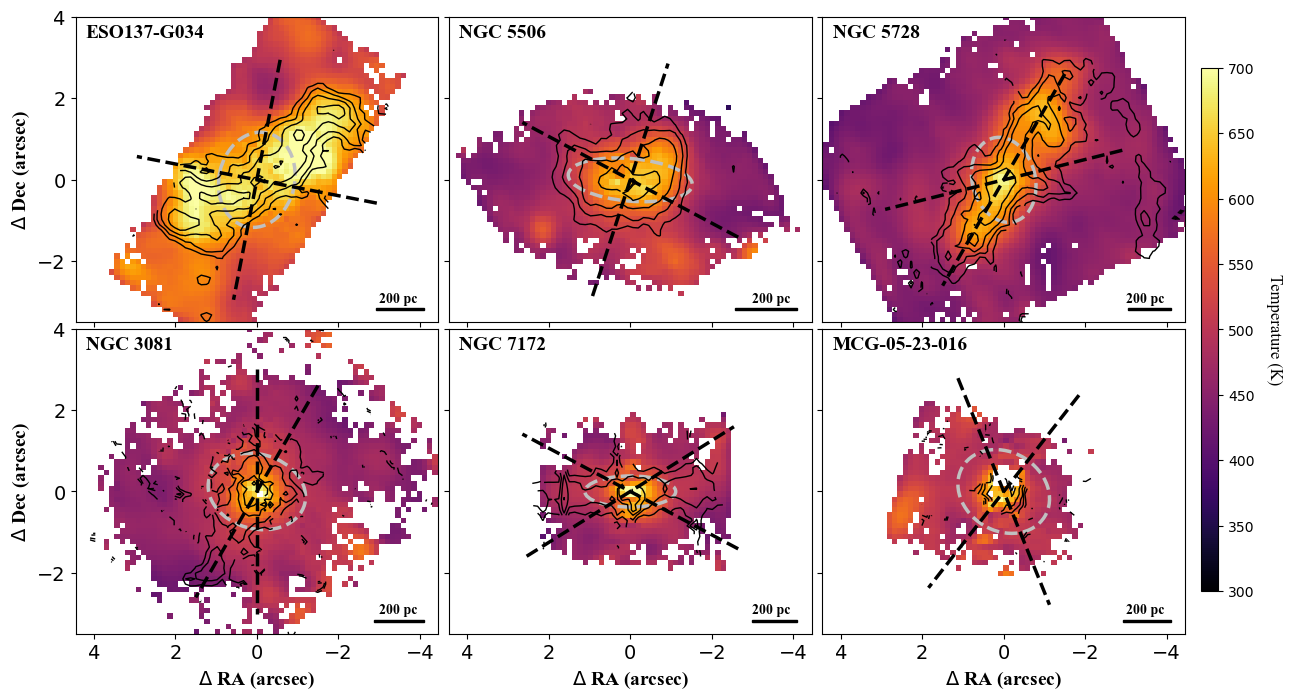}
\caption{$\rm H_2$ excitation temperature mapping for each object derived from the $\rm H_2$ 0-0 S(1)/S(5) line ratio. Panels are presented with estimated ionized mass outflow rate decreasing from top to bottom and left to right, with the highest ionized outflow rate in the top left (ESO 137-G034)  and lowest in (MCG-05-23-16) in the bottom right. The dashed black line indicates the edges of the ionization bicone (see Table \ref{Galaxy Fundamentals}). The gray ellipse represents a deprojected aperture of the central 400 pc (r = 200 pc). For MCG 05-23-016, since no information on opening angle is available, a default of 60$^{\circ}$ has been used. Black contours represent the [Fe~II]$_{5.34 \rm \mu m}$ flux contours ranging from 10$^{-16.45}$ and 10$^{-15.7}$ erg $\rm s^{-1}$\,$\rm cm^{-2}$  (0.25 dex steps) for ESO\,137-34, 10$^{-15.95}$ and 10$^{-15.2}$ erg $\rm s^{-1}$\,$\rm cm^{-2}$ (0.25 dex steps) for NGC\,5506, 10$^{-16.7}$ and 10$^{-15.95}$ erg $\rm s^{-1}$\,$\rm cm^{-2}$ (0.25 dex steps) for NGC\,5728, 10$^{-17.2}$ and 10$^{-16.45}$ erg $\rm s^{-1}$\,$\rm cm^{-2}$ (0.25 dex steps) for NGC\,3081,  10$^{-16.6}$ and 10$^{-15.85}$ erg $\rm s^{-1}$\,$\rm cm^{-2}$ (0.25 dex steps) for NGC\,7172, and 10$^{-17.2}$ and 10$^{-16.45}$ erg $\rm s^{-1}$\,$\rm cm^{-2}$ (0.25 dex steps) for MCG-0523-016. The gray dashed regions depict nuclear aperture investigated in Section \ref{sec:Central 400 pc}.     
\label{Excitation and Ionized Emission}}
\end{figure*}

Overlain on the temperature maps for each object  (Figure \ref{Excitation and Ionized Emission}), we include the emission contours for [Fe~II]$_{5.34 \mu m}$. Iron emission can result from photoionization, but is expected to be enhanced by shock induced dust destruction (\citealt{koo2016infrared, pereira2024extended, riffel2025impact}). Dust destruction resulting from shattering and sputtering in shocked environments is understood to release refractory elements such as iron that were previously trapped within dust grains \citep{jones1994grain, tielens1994physics, jones1996grain}. This association with shocks makes iron ionized emission of particular interest and we therefore utilize it trace potentially AGN driven excitation. 

Qualitatively speaking, the three objects in the sample with the highest ionized outflow rates (ESO137-G034, NGC\,5506, and NGC\,5728) show similar elongation and distribution of  $\rm H_2$ temperature and [Fe~II]$_{5.34 \mu m}$ emission (Figure \ref{Excitation and Ionized Emission}). The distribution of increased temperature and [Fe~II]$_{5.34 \mu m}$ emission in these targets is spatially coincident with the ionization cones of each respective target although the perceived alignment with the cone varies. In the case of NGC\,5728 and NGC\,5506, these features track along one or both of the edges of the ionization cone, tempting us to speculate that these correlations manifest where the ionization cone intersects the circumnuclear disk. For ESO\,137-G034, however, these extended features appear to bisect the opening angle of the ionization cone and reside primarily within the cone itself. It could be the case that these correlations exist independent of the relative geometry between the ionization cone and the circumnuclear disk, or this could be an artifact of our line of sight viewing of the ionization cone and the circumnuclear disk with intereaction between the cone and disk occuring on the far side of the cone. More detailed geometric modeling of the circumnuclear disk and ionziation cone is needed to fully assess this.

For NGC\,3081, NGC\,7172, and MCG-05-23-016, targets presenting lower ionized outflow rates, the temperature distribution is more centrally concentrated and [Fe~II]$_{5.34 \mu m}$ emission is not as clearly correlated with molecular gas temperature. We observe some extension of [Fe~II]$_{5.34 \mu m}$ emission for NGC\,3081 along the South-East portion of the ionization bicone, however this extension is not clearly observed in the $\rm H_2$ temperature distribution, potentially suggesting different excitation mechanisms. For NGC\,7172, we observe extension of [Fe~II]$_{5.34 \mu m}$ emission along the circumnuclear disk perpendicular to the direction of the ionization bicone. We interpret this emission to  likely result from the star forming ring present within the circumnuclear region of the galaxy \citep{munoz2024biconical}. 

\begin{figure*}[ht]
\plotone{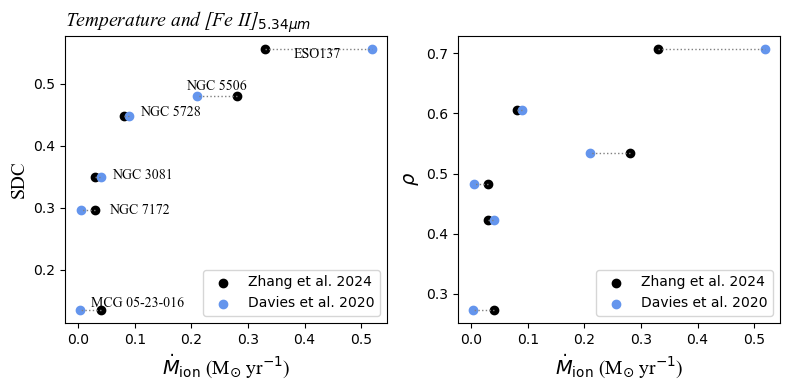}
\caption{Dice coefficient (left panel) and Pearson Coefficient (right panel) calculated from the $\rm H_2$ excitation map and [Fe~II]$_{5.34 \mu m}$ emission map  vs. the ionized outflow rate for each target.     
\label{Dice and Pearson v. Ionized Outflow Rate}}
\end{figure*}

To quantify the spatial correlation between the ionized [Fe~II]$_{5.34 \mu m}$ emission distribution and the molecular gas temperature, we calculate both the S\o rensen-Dice Coefficient (SDC), and the Pearson coefficient ($\rho$). SDC, which aims to quantify the spatial overlap between the two distributions, for two positive normalized arrays can be calculated as:
\begin{equation}
SDC = 1 - \frac{\sum\limits^i \sum\limits^j|Y_{i,j}-X_{i,j}|}{\sum\limits^i \sum\limits^j(|X_{i,j}| + |Y_{i,j}|)}
\label{Dice Coefficient Eqn}
\end{equation}
Where Y and X are the respective flux maps being compared. $\rho$, which quantifies the linear covariance, is computed as:
\begin{equation}
\rho = \frac{\sum\limits^i \sum\limits^j (X_{i,j} - \overline{X})(Y_{i,j} - \overline{Y})}{ \sqrt{\sum\limits^i \sum\limits^j(X_{i,j} - \overline{X})^2} \sqrt{\sum\limits^i \sum\limits^j(Y_{i,j} - \overline{Y})^2}}
\label{Pearson Coefficient Eqn}
\end{equation}
Where Y and X are the respective flux maps being compared, and $\overline{Y}$ and $\overline{X}$ are the average values of the flux map. Each of these quantities range from 0 to 1 with higher values indicating stronger levels of spatial overlap and covariance respectively. To ensure similar spatial scaling scales and resolution, the [Fe~II]$_{5.34 \mu m}$ emission map was convolved with the FWHM of the $\rm H_2$ 0-0 S(1) (17.0346 $\mu m$) PSF and we compute these values for 15 \arcsec{} nuclear apertures (indicated by gray square in Figure \ref{Fe II emission and Radio}. Additionally, to mitigate scaling effects, both maps were normalized by the maximum value of each respective map. For these calculations, any spaxels that were masked as non-detects were given a value of zero. Figure \ref{Dice and Pearson v. Ionized Outflow Rate} presents both the calculated SDC and $\rho$ as it relates to the ionized outflow rate. The strength of the spatial correlation between [Fe~II]$_{5.34 \mu m}$ and $\rm H_2$ temperature distribution, via the SDC and $\rho$, scales with ionized outflow rate.  The higher spatial correlation between [Fe~II]$_{5.34 \mu m}$ and temperature is suggestive of a shared excitation mechanism between molecular gas and iron. 

\subsection{[Fe~II]$_{5.34 \mu m}$ / [Ar II]$_{6.99 \mu m}$ Distributions} \label{subsec:[Fe II] and [Ar II]}

\begin{figure*}[ht]
\plotone{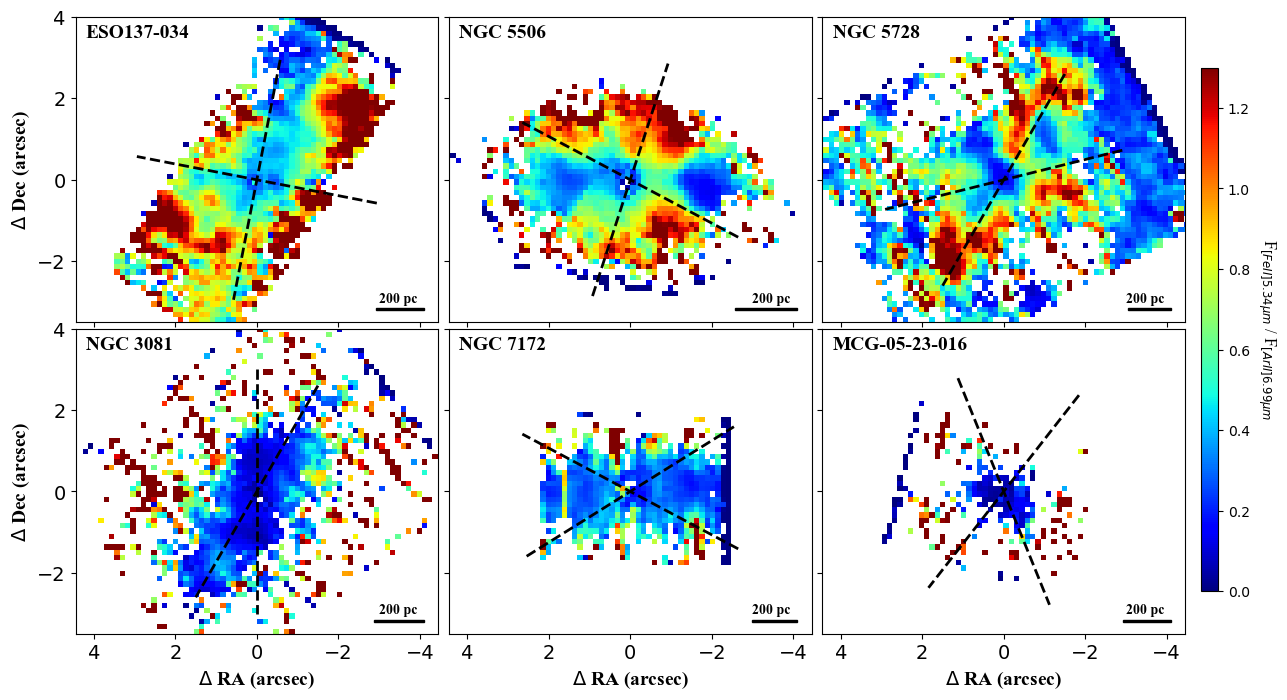}
\caption{[Fe~II]$_{5.34 \mu m}$ / [Ar II]$_{6.99 \mu m}$  map  vs. the ionized outflow rate for each target. Panels are presented with estimated ionized mass outflow rate increasing from top to bottom and left to right, with the highest ionized outflow rate in the top left (ESO 137-G034)  and lowest in (MCG-05-23-16) in the bottom right). Here, the [Fe~II]$_{5.34 \mu m}$ map has been convolved with the FWHM of the [Ar II]$_{6.99 
\mu m}$ PSF to match spatial scales. The dashed black line indicates the edges of the ionization bicone (see Table \ref{Galaxy Fundamentals}). For MCG 05-23-016, since no information on opening angle is available, a default of 60$^{\circ}$ has be used.   
\label{[Fe II] / [Ar II] Map}}
\end{figure*}

Of course, [Fe~II]$_{5.34 \mu m}$ emission is insufficient to conclude the presence of shocks as it can also result from photoionization and is only enhanced by dust destruction. Unlike iron which is depleted into dust \citep{Jenkins2009ApJ}, argon is a noble gas with a large photoization cross section and low condensation temperature with respect to iron (Ar: 48\,K; Fe: 1334\,K; see \citealt{lodders2003solar}). As such, it is not anticipated to suffer significant dust depletion \citep{Sofia1998, verma2003mid, amayo2021ionization} and will not be significantly enhanced by shock induced dust destruction making it a useful tracer for baseline photoionization (see \citealt{riffel2025impact}). To identify regions of potential shocks, we therefore look for enhancements of [Fe~II]$_{5.34 \mu m}$ with respect to [Ar II]$_{6.99 \mu m}$. Figure \ref{[Fe II] / [Ar II] Map} presents the [Fe~II]$_{5.34 \mu m}$/[Ar II]$_{6.99 \mu m}$ flux ratio maps for each target in the sample. For the three targets with the lowest ionized outflow rate (NGC\,3081, NGC\,7172, and MCG-05-23-016) we observe a relatively uniform distribution of [Fe~II]$_{5.34 \mu m}$/[Ar II]$_{6.99 \mu m}$. This is in stark contrast with the targets of the highest observed ionized outflow rates in which we observe regions of elevated [Fe~II]$_{5.34 \mu m}$ emission, particular in areas spatially coincident with the ionization bicone. 

Notably, regions with the most significant increase in [Fe~II]$_{5.34 \mu m}$/[Ar II]$_{6.99 \mu m}$ are at distance from the AGN rather than immediately surrounding the AGN. This may be a result of the immediate nuclear region being partially cleared of dust due to the harsh radiative field of the AGN which enhances both [Ar II]$_{6.99 \mu m}$ and [Fe~II]$_{5.34 \mu m}$ emission or perhaps an artifact of AGN driven photoionization being less significant at distance from the AGN. It could be that as the ionization cone or AGN outflows interact with the ambient interstellar medium (ISM), dust is destroyed and [Fe~II]$_{5.34 \mu m}$ is enhanced independent of [Ar II]$_{6.99 \mu m}$. As discussed in Section \ref{subsec:Temp and [Fe II]}, these regions of enhanced [Fe~II]$_{5.34 \mu m}$/[Ar II]$_{6.99 \mu m}$ may manifest where the ionization cone interacts with the curcumnuclear disk and modeling to constrain the relative geometries of the disk and outflow in each galaxies is needed to explore this further. Enhancement along the edge of the ionization cone rather than throughout the center of the cone where AGN photo-ionization is expected to be significant, as is particularly clear for NGC\,5728, supports this hypothesis and may be evidence that this enhancement results from shocks destroying dust at the interaction between the cone and disk. However, it is difficult to quantify the fraction of shock contribution to the observed [Fe II]$_{5.34 \mu m}$ emission. Recent theoretical models (\citealt{zhang2025theoretical}, see their Figure 8), find that in fast radiative shock environments [Fe~II]$_{5.34 \mu m}$/[Ar II]$_{6.99 \mu m}$ ratios of greater than 10  are predicted. While the sample galaxies display lower ratios throughout their circumnuclear regions, suggesting that AGN photo-ionization is the dominant excitation mechanism, it is still possible that fast radiative shocks are present and to some degree contribute to excitation and result in enhancing [Fe~II]$_{5.34 \mu m}$ emission. Alternatively, it could be that slower shocks driven by AGN winds or outflows may be responsible for the enhanced line ratio in these regions. In section \ref{subsec:Central 400pc [Fe II]/[Ar II]} we explore [Fe~II]$_{5.34 \mu m}$/[Ar II]$_{6.99 \mu m}$ for the nuclear region directly. 

\subsection{Radio Emission} \label{subsec:Radio and [Fe II]}

\begin{figure*}[ht]
\plotone{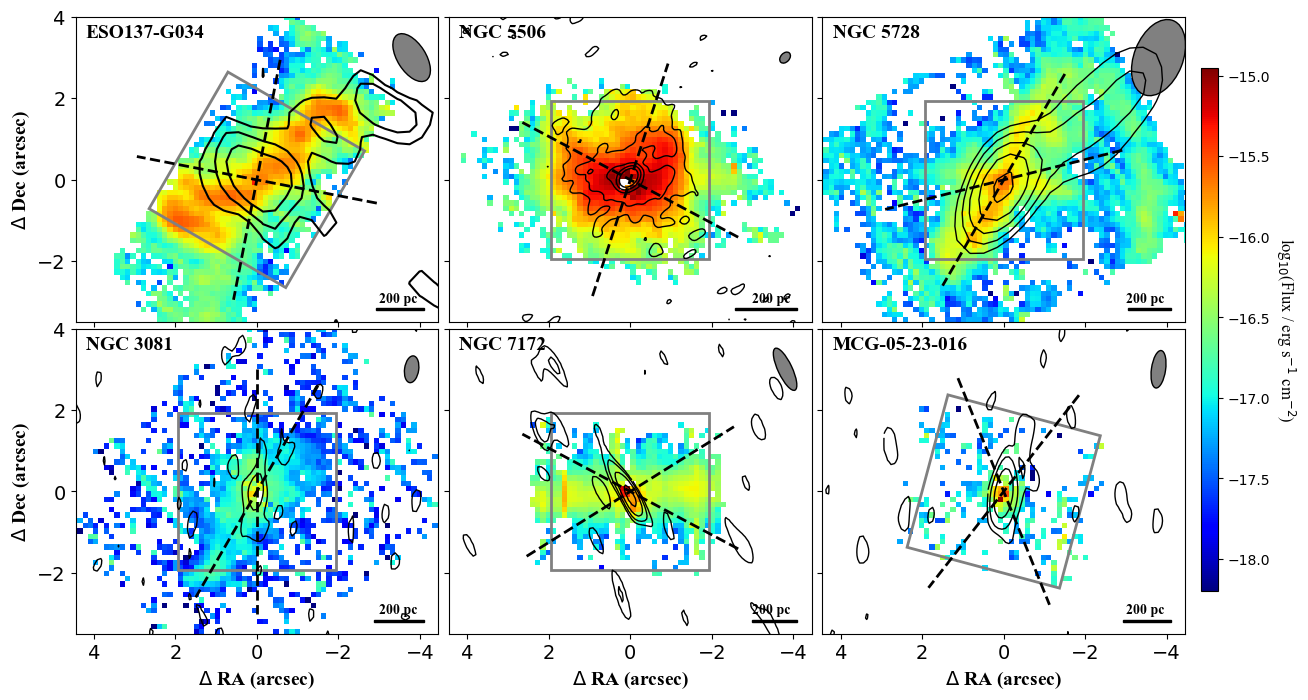}
\caption{[Fe~II]$_{5.34 \mu m}$ emission maps for each target. The black contours overlain on flux maps present the observed radio intensity for each object. The dashed black lines represent the edges of each targets ionization bicone. For MCG 05-23-016, since no information on opening angle is available, a default of 60$^{\circ}$ has be used. For ESO137-G034 the 8 GHz radio contour intervals \citep{morganti1999radio} range from 10$^{-3.5}$ to 10$^{-3.0}$ Jy/Beam with an 0.25 dex step. For NGC\,5506 the 8.46 GHz radio contour intervals \citep{schmitt2001jet} range from 10$^{-4.5}$ to 10$^{-2.0}$ Jy/Beam with an 0.5 dex step. For NGC\,5728 the 4.9 GHz radio contour intervals \citep{schmitt2001jet} range from 10$^{-3}$ to 10$^{-2.4}$ Jy/Beam with an 0.15 dex step. For NGC\,3081 the 4.9 GHz radio emission contour intervals \citep{NVAS} from 10$^{-4.0}$ to 10$^{-2.0}$ Jy/Beam with an 0.5 dex step. For NGC\,7172 the 4.9 GHz radio emission contour intervals \citep{NVAS} from 10$^{-3.75}$ to 10$^{-3.0}$ Jy/Beam with an 0.25 dex step. For MCG-05-23-016 the 4.9 GHz radio emission contour intervals \citep{NVAS} from 10$^{-3.75}$ to 10$^{-2.75}$ Jy/Beam with an 0.5 dex step. The gray elliipse in the top right corner of each panel indicates the respective beam size. The gray square aperture represents the region used for the computation of the S\o rensen-Dice and Pearson coefficient. 
\label{Fe II emission and Radio}}
\end{figure*}

Radio emission has been shown to be a tracer of AGN feedback and may correlate with shock excitation as jets impact ambient gas and drive AGN outflows  \citep{almeida2022diverse, audibert2023jet, de2023radio, holden2024alma, garcia2024galaxy}.  In fact, \cite{venturi2021magnum} showed the ionized gas mass and gas kinetic energy to correlate with jet power in four Seyfert AGN. In recent work, \cite{riffel2025blowing} concluded excess $\rm H_2$ emission in kinematically disturbed regions to be induced by shocks driven by radio jets interacting with the interstellar medium. Figure \ref{Fe II emission and Radio} presents the [Fe~II]$_{5.34 \mu m}$ emission distribution and radio intensity distribution for each target. For the three targets with the highest ionized outflow rates, we observe elongation of radio emission coincident with the distribution of [Fe~II]$_{5.34 \mu m}$ emission (and by extension the distribution of $\rm H_2$ excitation; Figure \ref{Excitation and Ionized Emission}). For ESO137-G034 and NGC\,5728, elongation of the jet is primarily along the north-western ionization cone and not does present the same extension along the south-eastern ionization cone. Whether it be AGN radio jets impacting and shocking circumnuclear gas and dust, or shocked gas from AGN winds/outflows ionizing and producing synchrotron radiation (e.g. \citealt{zakamska2014quasar, fischer2019dissection, fischer2023no}), we take this as further evidence that AGN activity plays an important role in molecular gas excitation in the three objects with the highest ionized outflow rates (ESO137-G034, NGC\,5506, and NGC\,5728). 

\section{The central 400 pc} \label{sec:Central 400 pc}

In an effort to apply uniform treatment over consistent spatial scales, we focus our analysis on the $\rm H_2$ within the central deprojected 400\,pcs (deprojected nuclear radii r\,$<$\,200 pc)  (Figure \ref{Excitation and Ionized Emission} gray apertures). This spatial scale was selected to ensure adequate signal-to-noise and to ensure a region at least as large as the PSF FWHM at the longest wavelength. To better constrain our analysis to the circumnuclear rotational disk, the nuclear apertures have been adjusted to match the inclination and position angle of the rotational disk circumnuclear to the AGN for each respective target (Table \ref{Galaxy Fundamentals}). For NGC\,3081 and ESO\,137-G034, the geometry of the circumnuclear region is not well constrained and thus we assume the circumnuclear disk at these scales matches that of the disk at larger scales reported in \cite{burtscher2021llama}. For NGC 5506 and NGC 7172, which are highly inclined (80$^{\circ}$ and 88$^{\circ}$, respectively), the aperture inclination has been limited to 70$^{\circ}$ to ensure the aperture is not smaller than the PSF FWHM. 

From the extracted integrated spectra (Appendix \ref{Appendix:Integrated Spectra and Population Levels} Figure \ref{Integrated Spectra}), we measure line fluxes for the $\rm H_2$ 0-0 S(1) through S(8) rotational emission lines visible to JWST MIRI/MRS and estimate the associated population levels ($\rm N_i / g_i$) following the methodology outlined in detail in \cite{delaney2025excitation}. Measured line fluxes for each target are presented in the top panel of Table \ref{Flux and LTE Fit Values}. Prior to computation of individual population levels, we have adjusted flux measurements for the extinction between rotational emission lines (details provided in Appendix \ref{Appendix:Extinction Corrections}). Assuming local thermodynamic equilibrium (LTE), where the warm $\rm H_2$ is expected to be primarily thermalized, we follow the procedure of \cite{togi2016lighting} in which the distribution of $\rm H_2$ density with temperature follows a single power law (SPL), such that 
\begin{equation}
dN \propto  T^{-\beta} dT 
\label{power law distribution}
\end{equation}
where $dN$ is the differential change in column density, and $\beta$ is the power law index which relates the proportions of warm to cool molecular gas mass.  We compare our measured population levels for the 8 measured rotational $\rm H_2$ emission lines to obtain the best fit free parameters for the LTE model ($N_{tot}$ and $\beta$) for the temperature range of 200-2000 $\rm K$. The lower limit of 200\,$\rm K$ is employed as for a thermalized gas at temperatures above this threshold, an Ortho-Para hydrogen ratio (OPR) of 3 can safely be assumed \citep{burton1992mid}. Immediately adjacent to the $H_2$ 0-0 S(7) emission line is the coronal emission line [Mg VII]$_{5.503 \mu m}$, which can be prevalent in the nuclear region around AGN. In some cases, the $H_2$ 0-0 S(7) emission is blended with [Mg VII]$_{5.503 \mu m}$ and difficult to de-blend, we therefore omit the S(7) associated population level from our model fit but still present the measured line fluxes and population level with the caveat of potential contamination.  

As Equation \ref{power law distribution} describes the particle column density as a function of temperature, we can obtain the population mean temperature from the LTE model best fit via
\begin{equation}
<T> =   \frac{\int_{T_l}^{T_u}  T \cdot T^{-\beta} dT}{\int_{T_l}^{T_u}   T^{-\beta} dT}
\label{Population Mean Temp Eqn}
\end{equation}
where $T_l$ = 200$\,K$ and $T_u$ = 2000$\,K$. For redundancy, we employ a second methodology and compute the isothermal gas temperature using the integrated $\rm H_2$ 0-0 S(1)/S(5) line flux ratio (Equation \ref{Excitation Temp Eqn}). Model best fit parameters and gas excitation temperatures are presented in Table \ref{Flux and LTE Fit Values}. Population levels and single power law best fit results are presented in Appendix \ref{Appendix:Integrated Spectra and Population Levels} Figure \ref{Best Fit LTE Models}.

\begin{deluxetable*}{cccccccc}
\tabletypesize{\scriptsize}
\tablewidth{0pt} 
\tablecaption{Measured $\rm H_2$ 0-0 S(1) through S(8), [Fe II]$_{5.34 \mu m}$, [Ar II]$_{6.99 \mu m}$, emission line flux from the 400\,pc nuclear apertures of each respective target where where a reliable line fit was achieved. The 1.4 GHz monochromatic radio luminosity of the 400\,pc nuclear aperture and estimated jet cavity power within this same aperture. $\rm H_2$ 0-0 S(1)/S(5) derived excitation temperature, LTE model best fit parameters as well as total estimate molecular mass within each aperture is also presented.   \label{Flux and LTE Fit Values}}
\tablehead{
\colhead{Emission Line} & \colhead{$\lambda_{rest}$ ($\rm \mu$m)} &  \multicolumn{6}{c}{Deprojected r $<$ 200\,pc Nuclear Line Flux ($\rm  10^{-16} erg \cdot s^{-1} cm^{-2}$)$^a$} \\
\colhead{} & \colhead{} & \colhead{ESO 137-G034} & \colhead{NGC 5506} & \colhead{NGC 5728} & \colhead{NGC 3081} & \colhead{NGC 7172} & \colhead{MCG-05-23-16}
} 
\colnumbers
\startdata 
\textbf{H$_2$ 0-0 S(1)} & 17.035 &  104.3  $\pm$   2.5 &  233.1  $\pm$   6.2 &  155.0  $\pm$   1.4 &  113.0  $\pm$   1.5 &  46.2  $\pm$   0.7 &  100.2  $\pm$   3.6 \\
\textbf{H$_2$ 0-0 S(2)} & 12.279 &  83.6  $\pm$   0.5 &  179.7  $\pm$   2.6 &  113.1  $\pm$   1.0 &  58.4  $\pm$   1.5 &  28.8  $\pm$   1.3 &  56.7  $\pm$   5.4 \\
\textbf{H$_2$ 0-0 S(3)} & 9.6649 &  275.6  $\pm$   0.6 &  243.6  $\pm$   1.9 &  194.3  $\pm$   1.6 &  160.9  $\pm$   0.6 &  20.9  $\pm$   0.4 &  130.3  $\pm$   1.3 \\
\textbf{H$_2$ 0-0 S(4)} & 8.0251 &  123.0  $\pm$   0.5 &  183.6  $\pm$   3.0 &  136.9  $\pm$   0.9 &  62.1  $\pm$   0.8 &  33.0  $\pm$   1.4 &  41.6  $\pm$   1.1 \\
\textbf{H$_2$ 0-0 S(5)} & 6.9095 &  286.9  $\pm$   1.5 &  498.7  $\pm$   3.3 &  300.6  $\pm$   1.7 &  139.0  $\pm$   1.6 &  71.4  $\pm$   1.6 &  100.3  $\pm$   2.3 \\
\textbf{H$_2$ 0-0 S(6)} & 6.1086 &  58.2  $\pm$   0.4 &  102.4  $\pm$   2.3 &  65.7  $\pm$   0.5 &  24.2  $\pm$   0.6 &  --$^b$ &  21.3  $\pm$   1.8 \\
\textbf{H$_2$ 0-0 S(7)$^c$} & 5.5112 &  107.5  $\pm$   5.7 &  300.2  $\pm$   4.0 &  229.8  $\pm$   3.8 &  90.8  $\pm$   3.6 &  41.5  $\pm$   1.1 &  56.4  $\pm$   0.0 \\
\textbf{H$_2$ 0-0 S(8)} & 5.053 &  32.7  $\pm$   0.3 &  69.9  $\pm$   4.4 &  36.3  $\pm$   0.5 &  20.4  $\pm$   1.0 &  9.7  $\pm$   1.0 &  13.8  $\pm$   1.2 \\
\textbf{[Fe II]} & 5.34 &  214.1  $\pm$   0.7 &  848.5  $\pm$   4.5 &  122.8  $\pm$   0.4 &  32.8  $\pm$   0.5 &  64.5  $\pm$   0.8 &  38.1  $\pm$   0.5 \\ \textbf{[Ar II]} & 6.985 &  486.2  $\pm$   1.7 &  2231.9  $\pm$   13.0 &  380.6  $\pm$   3.6 &  286.9  $\pm$   0.6 &  245.0  $\pm$   2.0 &  591.3  $\pm$   1.4 \\
\cline{1-8}
\multicolumn{8}{c}{\textbf{Radio Power}}  \\
\textbf{Log[L$_{1.4GHz}$ / W Hz$^{-1}$]}  &   &   22.3  &   22.8  &   22.9  &   21.6  &   22.0  &   22.6 \\
\textbf{Log[P$_{cav}$ / W Hz$^{-1}$]$^d$}  &   &   25.3  $\pm$   0.02  &   25.4  $\pm$   0.01  &   25.5  $\pm$   0.02  &   25.0  $\pm$   0.07  &   25.1  $\pm$   0.04  &   25.4  $\pm$   0.0001 \\
\cline{1-8}
\multicolumn{8}{c}{\textbf{S(5) / S(1) Derived Gas Temperature$^e$}}  \\
\textbf{Temperature (K)}  &   &   645.3  $\pm$   2.9  &   606.8  $\pm$   3.0  &   597.6  $\pm$   1.1  &   562.3  $\pm$   1.6  &   562.3  $\pm$   2.6  &   546.4  $\pm$   3.7 \\
\cline{1-8}
\multicolumn{8}{c}{\textbf{LTE Fit Parameters$^f$}}  \\
\textbf{$\beta$}  &   &   3.83  $\pm$   0.15  &   4.04  $\pm$   0.11  &   4.12  $\pm$   0.12  &   4.43  $\pm$   0.18  &   4.26  $\pm$   0.11  &   4.53  $\pm$   0.12 \\
\textbf{log[$N_{H_2}$ \ $\rm cm^{-2}$]}  &   &   20.7  $\pm$   0.1  &   21.2  $\pm$   0.1  &   21.1  $\pm$   0.1  &   20.7  $\pm$   0.1  &   20.9  $\pm$   0.1  &   20.6  $\pm$   0.06\\
\textbf{log[M$_{\rm H_2}$ / $M_{\odot}$]}  &   &   5.85 $^{ + 0.71 } _{ - 0.79 }$  &   6.07 $^{ + 0.86 } _{ - 0.92 }$  &   6.23 $^{ + 0.81 } _{ - 0.88 }$  &   5.88 $^{ + 0.66 } _{ - 0.75 }$  &   5.78 $^{ + 0.90 } _{ - 0.95 }$  &   5.85 $^{ + 0.86 } _{ - 0.92 }$ \\
\textbf{Average Temperature (K)}  &   &   305.3  $\pm$   7.0  &   295.8  $\pm$   4.5  &   292.5  $\pm$   4.6  &   281.4  $\pm$   5.2  &   287.2  $\pm$   3.9  &   278.3  $\pm$   3.3 \\
\enddata
\tablecomments{$^a$Flux values presented do not have extinction corrections applied. $^b$Due to a noisy continuum and potential blending with an oversampling wiggle pattern, a reliable fit was not achieved for this line. $^c$Due to the proximity of the $\rm H_2$ 0-0 S(7) emission line, there is potential for contamination of [Mg VII]$_{5.503 \mu m}$ emission. $^d$Jet cavity power within our deprojected 400\,pc nuclear apertures estimated via \cite{birzan2008radiative} equation 16. The scatter for the 1.4 GHz jet cavity power relation is 0.85 dex.  $^e$Excitation temperature is derived using extinction corrected line flux. Reported errors correspond to the standard deviation obtained from 1000 Monte Carlo realizations, incorporating the flux uncertainties. $^f$LTE modeling and associated parameters are derived using extinction corrected line flux. For LTE parameters, the error presented represents the fit error combined in quadrature the standard deviation obtained from 1000 Monte Carlo realizations incorporating the flux uncertainties.}
\end{deluxetable*}

\subsection{Excitation and Ionized Mass Outflow Rate} \label{subsec:Central 400pc Temperature}

For this analysis, we compare our observations to ionized mass outflow rates estimated using two different methodologies: \cite{davies2020ionized} who used the optical [O III] luminosity, and more recently, \cite{zhang2024galaxy} used the MIR line [Ne~V]$_{14.322 \mu m}$. Figure  \ref{Central 400 pc Figure - All plots} (top panels) presents the measured $\rm H_2$ gas temperature estimated via the S(1)/S(5) line ratio, the population mean temperature estimated from the SPL modeling results, and the estimated column density from the LTE modeling. Temperature estimates derived from the SPL-LTE modeling are lower than those from the S(1)/S(5) line ratio, as rather than assume a single gas temperature, the SPL-LTE modeling assumes a distribution of particles with most of the molecular mass occupying lower energy states (i.e. cooler gas). 

We observe a positive correlation between the ionized outflow rate and both the S(1)/S(5) derived gas temperature and the population mean temperature derived from the SPL-LTE model. This suggests that additional excitation energy is imparted onto the $\rm H_2$ of the circumnuclear region around AGN with higher ionized outflow rates. As targets within this sample are generally of similar bolometric luminosity (see Table \ref{Galaxy Fundamentals}), and with all things being equal, AGN photoionization is likely insufficient to explain this difference. It is possible that the strength of geometric coupling between the ionization cone and the circumnuclear disk (e.g. \citealt{almeida2022diverse}) may vary between objects, and targets with stronger coupling are more significantly impacted. Stronger disk-ionization cone coupling could result in more direct radiative warming of the molecular disk and have higher potential for outflow induced shocks. 

As a result these objects may present, higher gas temperatures, stronger correlations between $\rm H_2$ excitation and [Fe~II]$_{5.34 \mu m}$ emission, and more efficient mass loading thereby bolstering the ionized outflow rate. Estimated column densities of warm $\rm H_2$ (T $>$ 200\,K) from the SPL-LTE model fit  also appear to show some positive corelation with ionized outflow rate, with the exception of ESO 137-G034 (interestingly, the object with the lowest bolometric luminosity), which presents a lower column density. In the case of NGC\,5506 and NGC\,5728, this could be the result of the AGN warming cold molecular gas in the nuclear region resulting in an increased column density of warm/hot gas. An explanation consistent with the nuclear deficiency of cold molecular gas observed by \cite{garcia2021galaxy, garcia2024deciphering} for these objects. It is possible that in the case of ESO\,137-G034, that AGN activity has cleared the nuclear region of some material resulting in a lower overall column density of warm $\rm H_2$. 

Our interpretation that shock heating from outflow activity, scaling with the strength of outflow-disk coupling, is congruent with the results of previous studies which have provided evidence for AGN ionized outflow behavior to correlate with potential coupling. Notably, \cite{bergmann2012resolved} (and references therein) highlighted evidence that mass loading is necessary to explain the tendency of ionized outflow to be orders of magnitude larger than the mass accretion rate. \cite{muller2011outflows}, using a sample of seven Seyferts (type 1.5-2), showed an anti-correlation between the half-opening angle of the outflow and both the molecular gas mass in the central 60 pc (r $<$ 30\,pc) as well as an anti-correlation between the half-opening angle of the outflow and the maximum outflow velocity. A broader opening angle of the ionization cone may increase the likelihood of outflow-disk coupling and thereby entrain molecular gas in the outflow and decrease the outflow velocity. 

\subsection{[Fe~II]$_{5.34 \mu m}$ / [Ar II]$_{6.99 \mu m}$ and Ionized Mass Outflow Rate} \label{subsec:Central 400pc [Fe II]/[Ar II]}

As discussed in section \ref{subsec:Temp and [Fe II]}, enhancement of [Fe~II]$_{5.34 \mu m}$ emission relative to [Ar II]$_{6.99 \mu m}$ could identify the presence of potential shocks. Figure \ref{Central 400 pc Figure - All plots} (center panels) presents measured [Fe~II]$_{5.34 \mu m}$ / [Ar II]$_{6.99 \mu m}$ for our central nuclear apertures. We report enhancement of [Fe~II]$_{5.34 \mu m}$ / [Ar II]$_{6.99 \mu m}$ to correlate positively with the ionized outflow rate as well as the estimated molecular gas temperature of the nuclear region. One explanation for this enhancement of [Fe~II]$_{5.34 \mu m}$ could be that dust destruction from shocks induced by disk-cone coupling, promoting mass loading of outflows and providing additional excitation energy to molecular gas. However, as discussed in Section \ref{subsec:[Fe II] and [Ar II]}, this interpretation comes with the caveat that theoretical models (\citealt{zhang2025theoretical}) have suggested fast radiative shocks to yield higher ratios ([Fe~II]$_{5.34 \mu m}$/[Ar II]$_{6.99 \mu m}$ $>$ 10) than observed here. Our observations are therefore consistent with AGN photo-ionization as the dominant excitation mechanism. As a result, we suspect that photo-ionization is dominant excitation mechanism in the nuclear region of these Seyferts and that while the enhanced [Fe~II]$_{5.34 \mu m}$ may indicate shocked regions, this excitation mechanism is a less significant contributor. It is also possible that iron is somehow enhanced due to some alternative radiative process resulting in dust destruction. Regardless, the positive scaling observed between this line ratio and temperature, both derived from our SPL-LTE model and the $\rm H_2$ 0-0 S(1)/S(5) line ratio, leads us to suspect that the mechanism driving this enhancement is likely shared by the $\rm H_2$ excitation. 

\subsection{Jet Power and Ionized Mass Outflow Rate} \label{subsec:Central 400pc Jet Power}

\begin{figure*}[ht]
\plotone{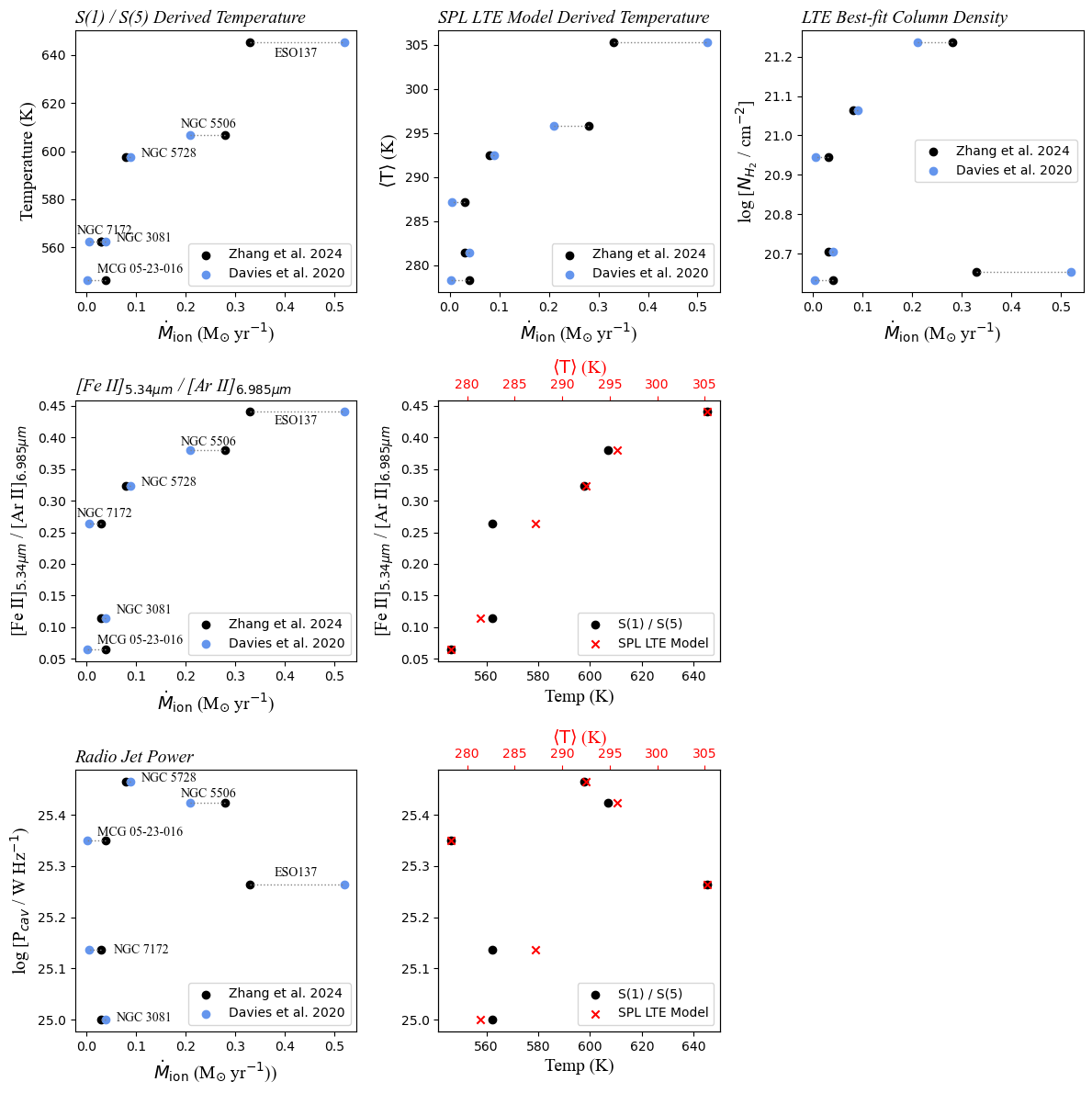}
\caption{Top Panels: From left to right: $\rm H_2$ 0-0 S(1)/S(5) derived temperature, LTE model derived temperature expectation value for the de-projected 400\,pc nuclear apertures, and the LTE model derived column density vs. ionized mass outflow rates estimated by \cite{davies2020ionized} and \cite{zhang2024galaxy}.
Center Panels: Left panel presents the measured line ratio [Fe~II]$_{5.34 \mu m}$ / [Ar II]$_{6.99 \mu m}$ for the de-projected 400\,pc nuclear apertures vs. ionized outflow rates estimated by \cite{davies2020ionized} and \cite{zhang2024galaxy}. Right panel depicts [Fe~II]$_{5.34 \mu m}$ / [Ar II]$_{6.99 \mu m}$ vs. the molecular gas temperature estimated from the LTE model (top axis) and derived from the $\rm H_2$ 0-0 S(5)/S(1) line ratio (bottom axis). 
Bottom Panel: Estimated radio jet cavity power vs. ionized outflow rate. 
Associated errors with quantities are included in Table \ref{Flux and LTE Fit Values}.
\label{Central 400 pc Figure - All plots}}
\end{figure*}

From the available radio data (Figure \ref{Fe II emission and Radio}) and assuming a power-law spectrum ($\rm S_{\nu}$ $\propto$ $\nu^{-\alpha}$; with $\alpha = 0.7$), we compute the 1.4 GHz monochromatic radio luminosity and the radio jet cavity power ($P_{cav}$, via \citealt{birzan2008radiative} equation 16) within the deprojected central 400\,pc region of each target (Table \ref{Flux and LTE Fit Values}). Figure \ref{Central 400 pc Figure - All plots} (bottom panels) depicts the estimated jet cavity power compared to the ionized outflow rate and estimated $\rm H_2$ temperature. We do not observe a strong correlation between the ionized outflow rate or the molecular gas temperature and estimated jet power in our sample. While there is degeneracy in the mechanisms warming molecular gas and potentially inducing shocks within the nuclear region (e.g. outflow components such as, winds, jets etc.),  we interpret this as evidence that while radio jets may be contributing to warming of the molecular gas and potentially inducing shocks, additional mechanisms are required to explain the correlation between the ionized mass outflow rate and both the molecular gas temperature and [Fe II]$_{5.34 \mu m}$/[Ar II]$_{6.99 \mu m}$ ratio of the nuclear region. We find the interpretation that ionization cone-disk coupling and on smaller scales shocks to be driving warming the nuclear molecular gas and promoting mass loading of the ionized outflow to be a plausible explanation. 

\section{Summary} \label{sec:Summary}

We have investigated the excitation of molecular gas within the deprojected central 400\,pc region (r\,$<$\,200 pc) as it relates to the observed ionized mass outflow rates estimated by \cite{davies2020ionized} and \cite{zhang2024galaxy}. Our analysis suggests the presence of shock excitation of the molecular gas to correlate with the observed ionized outflow rate. As part of this work, we report the following: 

\begin{itemize}
    \item     We observe the $H_2$ temperature distribution to be more strongly spatially correlated with [Fe~II]$_{5.34 \mu m}$ emission (a tracer of potential shocks) for the three targets in the sample with highest measured ionized outflow rates. We report the SDC and $\rho$ between the $\rm H_2$ temperature and [Fe~II]$_{5.34 \mu m}$ emission distribution to scale positively with the ionized outflow rate, and by extension nuclear molecular gas temperature.

    \item Based on SPL-LTE modeling of the S(1)-S(8) lines, we find the $\rm H_2$ temperature in the central 400\,pc to positively correlate with the ionized mass outflow rate: suggesting that higher excitation temperatures are associated with higher outflow rates. This observation is further corroborated by estimating the average gas excitation temperature via the $\rm H_2$ 0-0 S(1)/S(5) line ratio. 

    \item  We report observing enhanced [Fe~II]$_{5.34 \mu m}$ / [Ar II]$_{6.99 \mu m}$ in regions co-spatial with the ionization cone. In addition, we report the measured [Fe~II]$_{5.34 \mu m}$ / [Ar II]$_{6.99 \mu m}$  of the central 400\,pc (r\,$<$\,200 pc) nuclear region to scale positively with both ionized outflow rate and estimated molecular gas temperature, potentially indicative of shock induced dust destruction and $\rm H_2$ excitation. 

    \item We do not observe the estimated radio jet cavity power of the central 400\,pc (r\,$<$\,200 pc) nuclear region to strongly correlate with ionized mass outflow rate or nuclear molecular gas temperature of the sample.

\end{itemize}

We take the preceding points as evidence of AGN ionization cone and outflow interactions with the rotating molecular disk. The impact of coupled ionization cone and potentially outflow-induced shock heating appears to correlate with the ionized gas mass outflow rates. By extension, this may suggest that geometrical coupling between the ionization bicone and the circumnuclear disk scales with the outflow rate. This supports the interpretation that the observed ionized outflow rate is bolstered by mass loading, as targets with higher observed ionized outflow rates show evidence of outflow-disk interactions. We therefore propose that AGN feedback may well be a geometric problem to solve as much as an energetic one.

\section*{Acknowledgments}

We thank the referee for their insightful comments which have resulted in an improved manuscript. This research utilized the unique and powerful tools included within Astropy, a community developed python package \citep{robitaille2013astropy}. We thank Raffaella Morganti for kindly providing the 8GHz ATCA radio data for ESO137-G034. This work is based on observations with the NASA/ESA/CSA James Webb Space Telescope obtained from MAST at the Space Telescope Science Institute, which is operated by the Association of Universities for Research in Astronomy, Incorporated, under NASA contract NAS5- 03127. Support for Program number JWST- GO-01670 was provided through a grant from the STScI under NASA contract NAS5- 03127. D. D., L.Z., C.P., and E.K.S.H acknowledge grant support from the Space Telescope Science Institute (ID: JWST-GO-01670). D.D. and E. K. S. H. acknowledge support from the NASA Astrophysics Data Analysis Program (22-ADAP22-0173). M.P.S. acknowledges support under grants RYC2021-033094-I, CNS2023-145506, and PID2023-146667NB-I00 funded by MCIN/AEI/10.13039/501100011033 and the European Union NextGenerationEU/PRTR. A.A.H. and L.H.M. acknowledge support from grant PID2021-124665NB-I00 funded by MCIN/AEI/10.13039/501100011033 and by ERDF A way of making Europe. E.B. acknowledges support from the Spanish grants PID2022-138621NB-I00 and PID2021-123417OB-I00, funded by MCIN/AEI/10.13039/501100011033/FEDER, EU. C.R.A. and A.A. acknowledge support from the Agencia Estatal de Investigación of the Ministerio de Ciencia, Innovación y Universidades (MCIU/AEI) under the grant “Tracking active galactic nuclei feedback from parsec to kiloparsec scales”, with reference PID2022-141105NB-I00 and the European Regional Development Fund (ERDF). A.A. also acknowledges support from the European Union (WIDERA ExGal-Twin, GA 101158446). S.G.B and F.E. acknowledges support from the Spanish grant PID2022-138560NB-I00, funded by MCIN/AEI/10.13039/501100011033/FEDER, EU. I.G.B. is supported by the Programa Atracci\'on de Talento Investigador ``C\'esar Nombela'' via grant 2023-T1/TEC-29030 funded by the Community of Madrid. D.E.A. is supported by the "Becas Estancias Posdoctorales por México" EPM(1) 2024 (CVU: 592884) program of SECIHTI and acknowledges financial support from PAPIIT UNAM IN109123 and "Ciencia de Frontera" CONAHCyT CF2023-G100. S.F.H. acknowledges support through UK Research and Innovation (UKRI) under the UK government’s Horizon Europe Funding Guarantee (EP/Z533920/1, selected in the 2023 ERC Advanced Grant round) and an STFC Small Award (ST/Y001656/1). I.G.B. is supported by the Programa Atracci\'on de Talento Investigador ``C\'esar Nombela'' via grant 2023-T1/TEC-29030 funded by the Community of Madrid. C.R. acknowledges support from SNSF Consolidator grant F01\$-\$13252, Fondecyt Regular grant 1230345, ANID BASAL project FB210003 and the China-Chile joint research fund. R.A.R. acknowledges the support from the Conselho Nacional de Desenvolvimento Científico e Tecnológico (CNPq; Projects 303450/2022-3, and 403398/2023-1), the Coordenação de Aperfeiçoamento de Pessoal de Nível Superior (CAPES; Project 88887.894973/2023-00), and Fundação de Amparo à Pesquisa do Estado do Rio Grande do Sul (FAPERGS; Project 25/2551-0002765-9).


%

\vspace{5mm}
\facilities{JWST (MIRI)}


\software{astropy \citep{robitaille2013astropy}}



\appendix

\section{Appendix} \label{Appendix A}

\subsection{Evidence for AGN Wind Interactions} \label{subsec:Evidence_for_Winds}

Here we provide a brief overview of each target and potential evidence for coupling/AGN wind-disk interactions. 

\subsubsection{ESO 137-G034} \label{subsec:ESO137}

ESO 137-G034 is a relatively face on lenticular galaxy with a weak nuclear bar housing an obscured (Seyfert 2) AGN (NED\footnote{The NASA/IPAC Extragalactic Database (NED) is operated by the Jet Propulsion Laboratory, California Institute of Technology, under contract with the National Aeronautics and Space Administration.}). Of the GATOS cycle\,1 Sample, this object presents the largest observed ionized outflow rate of 0.52/0.33 $\rm M_{\odot}   yr^{-1}$ (\citealt{davies2020ionized}/\citealt{zhang2024galaxy}) and the lowest bolometric luminosity Log[L$_{\rm AGN}$ / erg s$^{-1}$] = 43.4 \citep{davies2020ionized}. \cite{zhang2024galaxy} identified highly kinematically disturbed regions (KDR) perpendicular to the ionization cone, potentially indicating shocked regions from AGN wind interactions. More recently, \cite{krol2026probing}, using X-ray observations from Chandra and spectral modeling showed that bicone emission likely results from a combination of photoionized gas and shocked plasma, suggesting AGN kinetic feedback.

\subsubsection{NGC 5506} \label{subsec:NGC5506}

NGC\,5506 is a peculiar edge on spiral galaxy housing an obscured (Seyfert type 1.9/1i\footnote{This object is classified as 'S1i' since it shows broad Paschen lines in the infrared, according to the nomenclature of \cite{veron2006catalogue}.} ) AGN \citep{garcia2021galaxy}. The ionization bicone of NGC 5506 is oriented from NE-SW and is nearly perpendicular to the disk of the galaxy. Both ionized and molecular (cool, warm, and hot) outflows have been identified in previous works (\citealt{davies2020ionized, riffel2023agnifs, zhang2024galaxy, esposito2024agn, delaney2025excitation}). The geometry of the ionization bicone and rotational disk was modeled by \cite{fischer2013determining} and a potential interaction between the ionization cone and disk was suggested by \cite{esposito2024agn}, who also identified cool molecular outflows using observations from the Atacama Large Millimeter Array (ALMA). Further, \cite{delaney2025excitation} using JWST Medium Resolution Spectroscopy (MRS) IFU data identified a region of warm molecular gas entrained in the AGN outflows and concluded AGN induced shocks are a likely excitation mechanism for the the nuclear molecular gas. Additionally, \cite{zhang2024galaxy} identified a KDR within the western portion of the disk perpendicular to the ionization bicone. 

\subsubsection{NGC 5728} \label{subsec:NGC5728}

NGC 5728 is a lenticular galaxy with a weak nuclear bar surrounded by a nuclear star forming ring and host an obscured (Seyfert 1.9) AGN (\citealt{shimizu2019multiphase}, NED). The geometry of the ionization cone, rotational disk, and multiphase gas kinematics were studied by \cite{shimizu2019multiphase} who reported holes in CO (2-1) emission of the rotating disk potentially suggesting gas removal, however found the structure of the circumnuclear region to be unaffected by the AGN. Expanding on this, \cite{davies2024gatos} used JWST/MRS IFU observations and to investigate $\rm H_2$ excitation within the central kpc and reported the warmest molecular gas to be present along the edges of the ionization cone where the molecular gas kinematics are significantly disturbed. \cite{davies2020ionized} argued that the ionization cone and rotational disk are coupled and molecular gas of the rotational disk is shocked and kinematically disturbed as it rotates through the ionization bicone of the AGN. This kinematic disturbance is consistent with the observations of \cite{zhang2024galaxy} who reported KDRs of ionized gas perpendicular to the direction of the ionization cone.

\subsubsection{NGC 3081} \label{subsec:NGC 3081}

NGC 3081 is a lenticular galaxy and is often characterized by it's three outer resonance rings and an inner resonance ring which encircles a weak nuclear bar and connects to two star forming arms \citep{ferruit2000hubble, buta1998ngc, buta2004hubble}. \cite{eguchi2011suzaku}. Housed within the center of this galaxy is an obscured (Seyfert 2) AGN which \cite{eguchi2011suzaku} argured resembeled a deeply burried AGN (shrouded in a very geometrically thick torus with a small opening angle; \citealt{ueda2007suzaku}). \cite{schnorr2016feeding} studied the gas kinematics of NGC 3081 using observations from Gemini Multi Object Spectrograph integral field spectrograph on the Gemini North telescope and presented some evidence for interaction between the disk and AGN outflows. More recently, \cite{delaney2025excitation} investigated the excitation of $\rm H_2$ using JWST/MRS IFU data and found little evidence for AGN driven shock excitation within the circumnuclear molecular gas. Circumnuclear dust, however, was studied by \cite{haidar2026gatos} of targets including NGC 3081 using JWST/MIRI imaging and found AGN illumination to be insufficient account for extended dust emission, indicating additional heating mechanisms such as shocks are required to explain observations.

\subsubsection{NGC 7172} \label{subsec:NGC7172}

NGC 7172 is an almost completely edge on ($\rm i_{disk} \approx $ 88$^{\circ}$) spiral galaxy with a heavily obscured (Seyfert 2) AGN. The nucleus of NGC 7172 is obscured both by a prominent dust lane \citep{smajic2012unveiling} and surrounded by a circumnuclear star forming ring extending up to 1.4 kpc which was identified by \cite{herrero2023agn} using CO(3-2) observations from the Atacama Large Millimeter/submillimeter Array (ALMA). \cite{herrero2023agn} also detected non-rotational motion of cold gas within the nuclear region indicating potentially outflowing of cold molecular gas. The ionization bicone of NGC 7172 is oriented nearly completely face on and presents a wide opening angle of $\approx$\,120$^{\circ}$ \cite{munoz2024biconical}. \cite{munoz2024biconical} investigated the ionized gas outflows and associated feedback in NGC 7172 indicated that the ionization cone is likely weakly coupled to the rotational disk and suggested that the reported ionized outflow rate may be underestimated as a result of extinction and projection effects.

\subsubsection{MCG-05-23-016} \label{subsec:MCG-05-23-016}

MCG-05-23-016 is a spiral galaxy hosting an obscured (Seyfert 2) AGN. It boasts the lowest ionized outflow rate of the GATOS Cycle\,1 sample. Despite having been shown to be a gas poor galaxy with little evidence of star formation activity and a relatively low stellar mass \cite{rosario2018llama}, the galaxy boasts the highest bolometric luminosity (Log[L$_{\rm AGN}$ / erg s$^{-1}$] = 44.3; \citealt{davies2020ionized}) of the GATOS Cycle\,1 sample. While evidence for geometric coupling between the ionization bicone and the rotaitonal disk or other wind interactions is limited, \cite{rosario2018llama} noted that the SED for this target is consistent with dust heated purely by the AGN. MCG-05-23-016 has been shown to contain a relatively weak compact radio ject \citep{orienti2010radio}, however \cite{esparza2025molecular} found this not to be connected with the kinematics of $\rm H_2$ gas.

\subsection{Emission Line Fitting and Map Generation} \label{Appendix:Line Fitting}

The four channels of JWST MIRI/MRS generate observations with differing scales and spatial resolutions. To compensate, in our analysis we have resampled all datacubes to match the smallest pixel grid across the channels (0.13\arcsec{}). The line fitting routine employed is similar to that described in \cite{delaney2025excitation} in which a 3rd-order Chebyshev polynomial fit was used to fit and subtract the local continuum and up to two gaussian components were used to fit emission lines. For error analysis, for each fit, 100 noise perturbed mock spectra were generated and the line was refit for each mock spectra, and the standard deviation of the fits were recorded as error. Line maps have been masked such that any spaxel with flux that did not meet the signal-to-noise threshold of 3 were masked. In addition, spaxels with measured velocity  or $|\sigma|$ $>$ 800 km/s have been masked. Where spaxel-by-spaxel analysis were performed (i.e. temperature mapping and estimation of Dice Coefficient and $\rho$), we have convolved the $\rm H_2$ 0-0 S(5) (6.9095 $\mu m$) and [Fe~II]$_{5.34 \mu m}$ line maps with the FWHM of the larger PSF of the $\rm H_2$ 0-0 S(1) (17.0346 $\mu m$) emission to match dataset resolution.

\subsection{Integrated Spectra and Population Levels} \label{Appendix:Integrated Spectra and Population Levels}

\begin{figure*}[ht]
\plotone{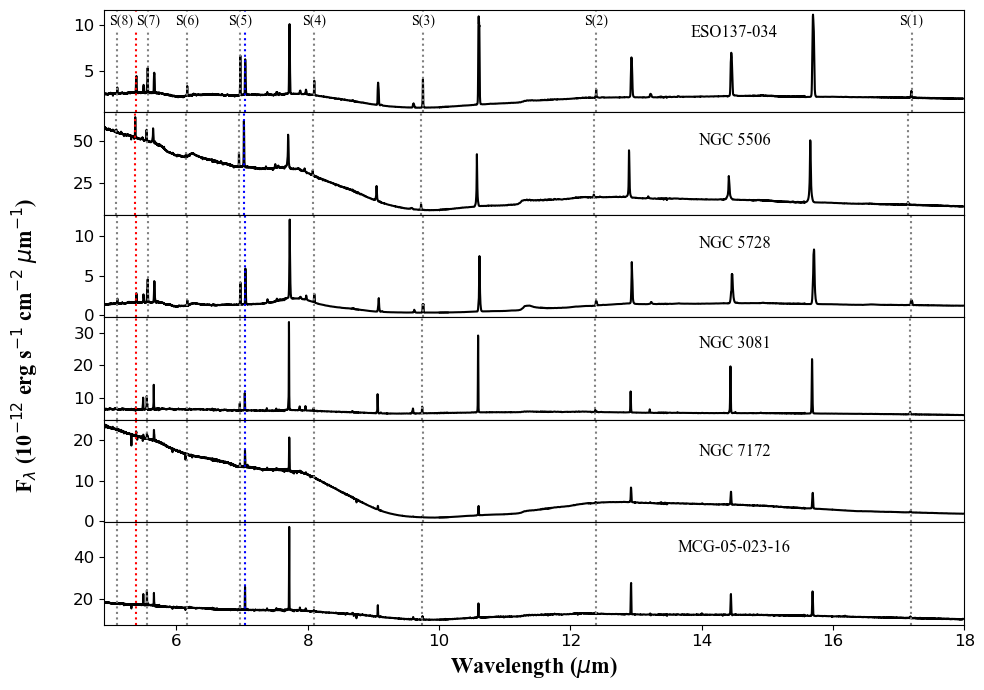}
\caption{Integrated spectra extracted from the 400\,pc nuclear aperture for each object at rest frame wavelength. The S(1) through S(8) rotational $\rm H_2$ emission lines are indentified by the black dotted line. Red dotted line identifies the position of the [Fe~II]$_{5.34 \mu m}$ emission line and the blue dotted line identifies the [Ar II]$_{6.99 \mu m}$ emisson line.      
\label{Integrated Spectra}}
\end{figure*}

\begin{figure*}[ht]
\plotone{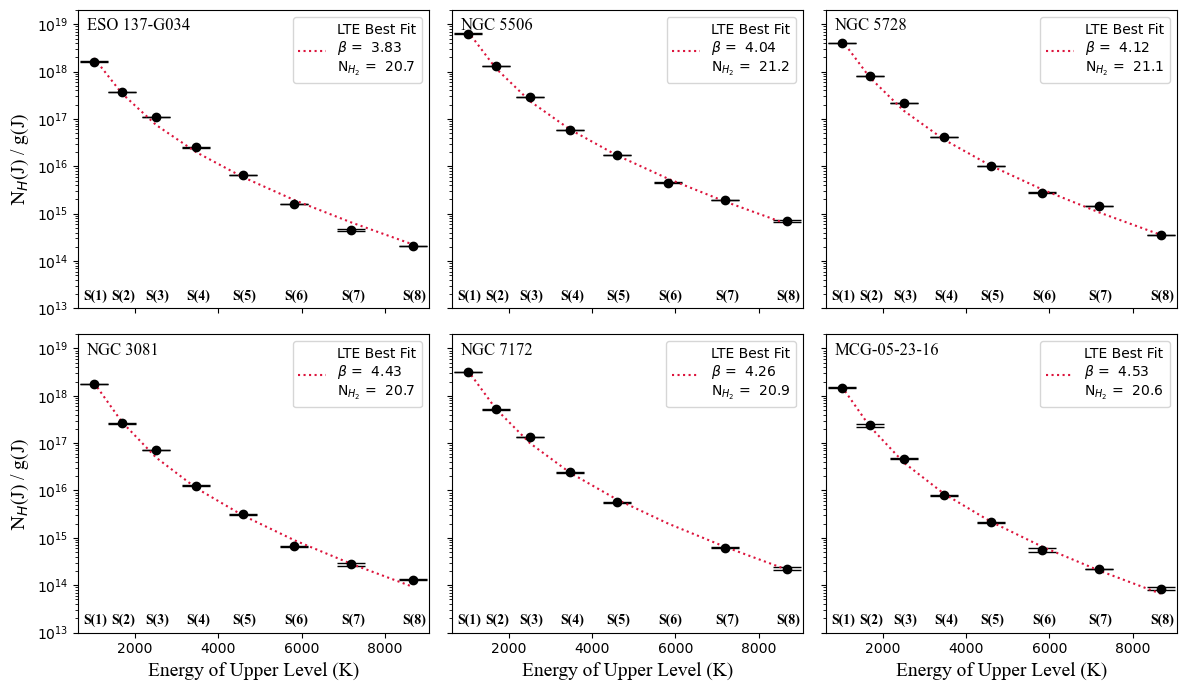}
\caption{Best fit LTE models for the S(1) through S(8) population levels. The S(7) population level was omitted from the model fit due to potential contamination by [Mg VII]$_{5.503 \mu m}$ emission.
\label{Best Fit LTE Models}}
\end{figure*}

Here we present the integrated spectra (Figure \ref{Integrated Spectra}) from the de-projected 400\,pc nuclear apertures and the resulting population levels with the best fit single power law model (Figure \ref{Best Fit LTE Models}) for each target. Nuclear apertures are  3.44 sq. \arcsec{} for ESO137-G034, 2.51 sq. \arcsec{} for NGC\,5506, 2.57 sq. \arcsec{} for NGC\,5728, 3.49 sq. \arcsec{} for NGC\,3081, 1.34 sq. \arcsec{} for NGC\,7172, and 3.58 sq. \arcsec{} for MCG-05-23-016. 

\subsection{Extinction Corrections} \label{Appendix:Extinction Corrections}

Extinction corrections were applied using the \cite{fritz2011line} infrared extinction curve which was derived for the Galactic center using hydrogen emission lines. Extinction was estimated using the adjacent S(3) and S(4) emission line flux relative to the intrinsic line flux predicted assuming LTE, and a Boltzmann's distribution at the temperature estimated via equation \ref{Excitation Temp Eqn}. As the S(1) and S(5) emission experience differential absorption we calculate the gas temperature with the extinction corrected flux and re-iterate this procedure until the computed temperature converges. Across the sample, differential reddening from the $\rm H_2$ 0-0 S(1) through S(8) lines is generally minor with NGC\,7172 being the most significantly impacted by extinction effects. For NGC\,3081, ESO137-G034, and MCG-05-23-16. For ESO137-G034 we estimate a differential reddening of 0.1 mag from the S(1) to S(8) emission. For NGC\,5506 and NGC\,5728 the differential reddening of 0.2 mag and for NGC\,7172 a differential reddening of 0.3 mag.  The S(3) emission line is impacted significantly by the 9.7\,$\mu m$ silicate absorption feature. The ratio $\rm F_{S(3)} / F_{S(1)}$ is adjusted by a factor of 1.5 for ESO137-G034, 2.5 for NGC\,5506, 2.4 for NGC\,5728, 1.6 for NGC\,3081, 5.3 for NGC\,7172, and 1.4 for MCG-05-23-016. We spot check our extinction correction via the S(3) associated population level (Figure \ref{Best Fit LTE Models}). As the resultant population level appears reasonable, we conclude that our extinction correction is valid.


\bibliography{ApJ_Letter_References}{}

@article{hickox2018,
  title={Obscured active galactic nuclei},
  author={Hickox, Ryan C and Alexander, David M},
  journal={Annual Review of Astronomy and Astrophysics},
  volume={56},
  pages={625--671},
  year={2018},
  publisher={Annual Reviews}
}

@article{wada2012radiation,
  title={Radiation-driven fountain and origin of torus around active galactic nuclei},
  author={Wada, Keiichi},
  journal={The Astrophysical Journal},
  volume={758},
  number={1},
  pages={66},
  year={2012},
  publisher={IOP Publishing}
}

@article{veilleux2005galactic,
  title={Galactic winds},
  author={Veilleux, Sylvain and Cecil, Gerald and Bland-Hawthorn, Joss},
  journal={Annu. Rev. Astron. Astrophys.},
  volume={43},
  number={1},
  pages={769--826},
  year={2005},
  publisher={Annual Reviews}
}

@article{veilleux2013fast,
  title={Fast molecular outflows in luminous galaxy mergers: evidence for quasar feedback from Herschel},
  author={Veilleux, S and Mel{\'e}ndez, M and Sturm, E and Gracia-Carpio, J and Fischer, J and Gonz{\'a}lez-Alfonso, E and Contursi, A and Lutz, D and Poglitsch, A and Davies, R and others},
  journal={The Astrophysical Journal},
  volume={776},
  number={1},
  pages={27},
  year={2013},
  publisher={IOP Publishing}
}

@article{almeida2017nuclear,
  title={Nuclear obscuration in active galactic nuclei},
  author={Ramos Almeida, Cristina and Ricci, Claudio},
  journal={Nature Astronomy},
  volume={1},
  number={10},
  pages={679--689},
  year={2017},
  publisher={Nature Publishing Group UK London}
}

@article{molina2023enhanced,
  title={Enhanced Star Formation Efficiency in the Central Regions of Nearby Quasar Hosts},
  author={Molina, Juan and Ho, Luis C and Wang, Ran and Shangguan, Jinyi and Bauer, Franz E and Treister, Ezequiel},
  journal={The Astrophysical Journal},
  volume={944},
  number={1},
  pages={30},
  year={2023},
  publisher={IOP Publishing}
}

@article{baumgartner201370,
  title={The 70 month Swift-BAT all-sky hard X-ray survey},
  author={Baumgartner, WH and Tueller, J and Markwardt, CB and Skinner, GK and Barthelmy, S and Mushotzky, RF and Evans, PA and Gehrels, N},
  journal={The Astrophysical Journal Supplement Series},
  volume={207},
  number={2},
  pages={19},
  year={2013},
  publisher={IOP Publishing}
}

@article{garcia2021galaxy,
  title={The Galaxy Activity, Torus, and Outflow Survey (GATOS)-I. ALMA images of dusty molecular tori in Seyfert galaxies},
  author={Garc{\'\i}a-Burillo, S and Alonso-Herrero, A and Almeida, C Ramos and Gonz{\'a}lez-Mart{\'\i}n, O and Combes, F and Usero, A and H{\"o}nig, S and Querejeta, M and Hicks, EKS and Hunt, Leslie Kipp and others},
  journal={Astronomy \& Astrophysics},
  volume={652},
  pages={A98},
  year={2021},
  publisher={EDP Sciences}
}

@article{esposito2024agn,
  title={AGN feedback in the Local Universe: Multiphase outflow of the Seyfert galaxy NGC 5506},
  author={Esposito, Federico and Alonso-Herrero, Almudena and Garc{\'\i}a-Burillo, Santiago and Casasola, Viviana and Combes, Fran{\c{c}}oise and Dallacasa, Daniele and Davies, Richard and Garc{\'\i}a-Bernete, Ismael and Garc{\'\i}a-Lorenzo, Bego{\~n}a and Mu{\~n}oz, Laura Hermosa and others},
  journal={Astronomy \& Astrophysics},
  volume={686},
  pages={A46},
  year={2024},
  publisher={EDP Sciences}
}

@article{fischer2013determining,
  title={Determining inclinations of active galactic nuclei via their narrow-line region kinematics. I. observational results},
  author={Fischer, TC and Crenshaw, DM and Kraemer, SB and Schmitt, HR},
  journal={The Astrophysical Journal Supplement Series},
  volume={209},
  number={1},
  pages={1},
  year={2013},
  publisher={IOP Publishing}
}

@article{davies2020ionized,
  title={Ionized outflows in local luminous AGN: what are the real densities and outflow rates?},
  author={Davies, R and Baron, Dalya and Shimizu, Taro and Netzer, Hagai and Burtscher, Leonard and De Zeeuw, P Tim and Genzel, Reinhard and Hicks, Erin KS and Koss, Michael and Lin, Ming-Yi and others},
  journal={Monthly Notices of the Royal Astronomical Society},
  volume={498},
  number={3},
  pages={4150--4177},
  year={2020},
  publisher={Oxford University Press}
}

@article{schnorr2016feeding,
  title={Feeding and feedback in NGC 3081},
  author={Schnorr-M{\"u}ller, Allan and Storchi-Bergmann, Thaisa and Robinson, Andrew and Lena, Davide and Nagar, Neil M},
  journal={Monthly Notices of the Royal Astronomical Society},
  volume={457},
  number={1},
  pages={972--985},
  year={2016},
  publisher={Oxford University Press}
}

@article{caglar2020llama,
  title={LLAMA: The MBH--$\sigma$⋆ relation of the most luminous local AGNs},
  author={Caglar, Turgay and Burtscher, Leonard and Brandl, Bernhard and Brinchmann, Jarle and Davies, Richard I and Hicks, Erin KS and Koss, Michael and Lin, Ming-Yi and Maciejewski, Witold and M{\"u}ller-S{\'a}nchez, Francisco and others},
  journal={Astronomy \& Astrophysics},
  volume={634},
  pages={A114},
  year={2020},
  publisher={EDP Sciences}
}

@article{riffel2023agnifs,
  title={The AGNIFS survey: spatially resolved observations of hot molecular and ionized outflows in nearby active galaxies},
  author={Riffel, Rogemar Andr{\'e} and Storchi-Bergmann, Thaisa and Riffel, Rog{\'e}rio and Bianchin, Marina and Zakamska, Nadia L and Ruschel-Dutra, D and Bentz, Misty Cherie and Burtscher, Leonard and Crenshaw, Daniel Michael and Dahmer-Hahn, LG and others},
  journal={Monthly Notices of the Royal Astronomical Society},
  volume={521},
  number={2},
  pages={1832--1848},
  year={2023},
  publisher={Oxford University Press}
}

@article{morganti1999radio,
  title={Radio continuum morphology of southern Seyfert galaxies},
  author={Morganti, R and Tsvetanov, ZI and Gallimore, J and Allen, MG},
  journal={Astronomy and Astrophysics Supplement Series},
  volume={137},
  number={3},
  pages={457--471},
  year={1999},
  publisher={EDP Sciences}
}

@article{buta1998ngc,
  title={NGC 3081: Surface photometry and kinematics of a classic resonance ring barred galaxy},
  author={Buta, R and Purcell, Guy B},
  journal={The Astronomical Journal},
  volume={115},
  number={2},
  pages={484},
  year={1998},
  publisher={IOP Publishing}
}

@article{buta2004hubble,
  title={A Hubble Space Telescope study of star formation in the inner resonance ring of NGC 3081},
  author={Buta, Ronald J and Byrd, Gene G and Freeman, Tarsh},
  journal={The Astronomical Journal},
  volume={127},
  number={4},
  pages={1982},
  year={2004},
  publisher={IOP Publishing}
}

@article{ferruit2000hubble,
  title={Hubble Space Telescope WFPC2 imaging of a sample of early-type Seyfert galaxies},
  author={Ferruit, Pierre and Wilson, Andrew S and Mulchaey, John},
  journal={The Astrophysical Journal Supplement Series},
  volume={128},
  number={1},
  pages={139},
  year={2000},
  publisher={IOP Publishing}
}

@article{wells2015mid,
  title={The mid-infrared instrument for the james webb space telescope, vi: The medium resolution spectrometer},
  author={Wells, Martyn and Pel, J-W and Glasse, Alistair and Wright, GS and Aitink-Kroes, Gabby and Azzollini, Ruym{\'a}n and Beard, Steven and Brandl, BR and Gallie, Angus and Geers, VC and others},
  journal={Publications of the Astronomical Society of the Pacific},
  volume={127},
  number={953},
  pages={646},
  year={2015},
  publisher={IOP Publishing}
}

@article{argyriou2023jwst,
  title={JWST MIRI flight performance: the medium-resolution spectrometer},
  author={Argyriou, Ioannis and Glasse, Alistair and Law, David R and Labiano, Alvaro and {\'A}lvarez-M{\'a}rquez, Javier and Patapis, Polychronis and Kavanagh, Patrick J and Gasman, Danny and Mueller, Michael and Larson, Kirsten and others},
  journal={Astronomy \& Astrophysics},
  volume={675},
  pages={A111},
  year={2023},
  publisher={EDP Sciences}
}

@article{togi2016lighting,
  title={Lighting the dark molecular gas: H2 as a direct tracer},
  author={Togi, Aditya and Smith, JDT},
  journal={The Astrophysical Journal},
  volume={830},
  number={1},
  pages={18},
  year={2016},
  publisher={IOP Publishing}
}

@article{pereira2014warm,
  title={Warm molecular gas temperature distribution in six local infrared bright Seyfert galaxies},
  author={Pereira-Santaella, Miguel and Spinoglio, Luigi and van der Werf, Paul P and L{\'o}pez, Javier Piqueras},
  journal={Astronomy \& Astrophysics},
  volume={566},
  pages={A49},
  year={2014},
  publisher={EDP Sciences}
}

@article{davies2024gatos,
  title={GATOS: missing molecular gas in the outflow of NGC 5728 revealed by JWST},
  author={Davies, R and Shimizu, T and Pereira-Santaella, M and Alonso-Herrero, A and Audibert, A and Bellocchi, E and Boorman, P and Campbell, S and Cao, Y and Combes, F and others},
  journal={Astronomy \& Astrophysics},
  volume={689},
  pages={A263},
  year={2024},
  publisher={EDP Sciences}
}

@article{herrero2023agn,
  title={AGN feedback in action in the molecular gas ring of the Seyfert galaxy NGC 7172},
  author={Alonso-Herrero, A Alonso and Garc{\'\i}a-Burillo, S and Pereira-Santaella, Miguel and Shimizu, T and Combes, F and Hicks, EKS and Davies, R and Almeida, C Ramos and Garc{\'\i}a-Bernete, I and H{\"o}nig, SF and others},
  journal={Astronomy \& Astrophysics},
  volume={675},
  pages={A88},
  year={2023},
  publisher={EDP Sciences}
}

@article{fritz2011line,
  title={Line derived infrared extinction toward the galactic center},
  author={Fritz, Tobias K and Gillessen, Stefan and Dodds-Eden, Katie and Lutz, Dieter and Genzel, Reinhard and Raab, Walfried and Ott, Thomas and Pfuhl, Oliver and Eisenhauer, Frank and Yusef-Zadeh, Farhad},
  journal={The Astrophysical Journal},
  volume={737},
  number={2},
  pages={73},
  year={2011},
  publisher={IOP Publishing}
}

@article{burton1992mid,
  title={Mid-infrared rotational line emission from interstellar molecular hydrogen},
  author={Burton, Michael G and Hollenbach, DJ and Tielens, AGG},
  journal={Astrophysical Journal, Part 1 (ISSN 0004-637X), vol. 399, no. 2, p. 563-572.},
  volume={399},
  pages={563--572},
  year={1992}
}

@article{pereira2024extended,
  title={Extended high-ionization [Mg IV] emission tracing widespread shocks in starbursts seen by JWST/NIRSpec},
  author={Pereira-Santaella, Miguel and Garc{\'\i}a-Bernete, Ismael and Gonz{\'a}lez-Alfonso, Eduardo and Alonso-Herrero, Almudena and Colina, Luis and Garc{\'\i}a-Burillo, Santiago and Rigopoulou, Dimitra and Arribas, Santiago and Perna, Michele},
  journal={Astronomy \& Astrophysics},
  volume={685},
  pages={L13},
  year={2024},
  publisher={EDP Sciences}
}

@article{koo2016infrared,
  title={Infrared [Fe II] emission lines from radiative atomic shocks},
  author={Koo, Bon-Chul and Raymond, John C and Kim, Hyun-Jeong},
  journal={arXiv preprint arXiv:1604.00748},
  year={2016}
}

@article{alonso2019nuclear,
  title={Nuclear molecular outflow in the Seyfert galaxy NGC 3227},
  author={Alonso-Herrero, A and Garc{\'\i}a-Burillo, S and Pereira-Santaella, M and Davies, RI and Combes, F and Vestergaard, M and Raimundo, SI and Bunker, A and D{\'\i}az-Santos, T and Gandhi, P and others},
  journal={Astronomy \& Astrophysics},
  volume={628},
  pages={A65},
  year={2019},
  publisher={EDP Sciences}
}

@article{davies2014fueling,
  title={Fueling active galactic nuclei. II. Spatially resolved molecular inflows and outflows},
  author={Davies, RI and Maciejewski, W and Hicks, EKS and Emsellem, E and Erwin, P and Burtscher, L and Dumas, G and Lin, M and Malkan, MA and M{\"u}ller-S{\'a}nchez, F and others},
  journal={The Astrophysical Journal},
  volume={792},
  number={2},
  pages={101},
  year={2014},
  publisher={IOP Publishing}
}

@article{cicone2014massive,
  title={Massive molecular outflows and evidence for AGN feedback from CO observations},
  author={Cicone, CLAUDIA and Maiolino, R and Sturm, E and Graci{\'a}-Carpio, J and Feruglio, Chiara and Neri, R and Aalto, Susanne and Davies, R and Fiore, Fabrizio and Fischer, J and others},
  journal={Astronomy \& Astrophysics},
  volume={562},
  pages={A21},
  year={2014},
  publisher={EDP Sciences}
}

@article{fabian2012observational,
  title={Observational evidence of active galactic nuclei feedback},
  author={Fabian, Andrew C},
  journal={Annual Review of Astronomy and Astrophysics},
  volume={50},
  number={1},
  pages={455--489},
  year={2012},
  publisher={Annual Reviews}
}

@article{ellison2018star,
  title={Star formation is boosted (and quenched) from the inside-out: radial star formation profiles from MaNGA},
  author={Ellison, Sara L and S{\'a}nchez, Sebastian F and Ibarra-Medel, Hector and Antonio, Braulio and Mendel, J Trevor and Barrera-Ballesteros, Jorge},
  journal={Monthly Notices of the Royal Astronomical Society},
  volume={474},
  number={2},
  pages={2039--2054},
  year={2018},
  publisher={Oxford University Press}
}

@article{esquej2013nuclear,
  title={Nuclear star formation activity and black hole accretion in nearby Seyfert galaxies},
  author={Esquej, Pilar and Alonso-Herrero, Almudena and Gonz{\'a}lez-Mart{\'\i}n, Omaira and H{\"o}nig, Sebastian F and Hern{\'a}n-Caballero, Antonio and Roche, P and Almeida, C Ramos and Mason, Rachel E and D{\'\i}az-Santos, Tanio and Levenson, Nancy A and others},
  journal={The Astrophysical Journal},
  volume={780},
  number={1},
  pages={86},
  year={2013},
  publisher={IOP Publishing}
}

@article{davies2007close,
  title={A close look at star formation around active galactic nuclei},
  author={Davies, RI and S{\'a}nchez, F M{\"u}ller and Genzel, R and Tacconi, LJ and Hicks, Erin KS and Friedrich, S and Sternberg, A},
  journal={The Astrophysical Journal},
  volume={671},
  number={2},
  pages={1388},
  year={2007},
  publisher={IOP Publishing}
}

@article{robitaille2013astropy,
  title={Astropy: A community Python package for astronomy},
  author={Robitaille, Thomas P and Tollerud, Erik J and Greenfield, Perry and Droettboom, Michael and Bray, Erik and Aldcroft, Tom and Davis, Matt and Ginsburg, Adam and Price-Whelan, Adrian M and Kerzendorf, Wolfgang E and others},
  journal={Astronomy \& Astrophysics},
  volume={558},
  pages={A33},
  year={2013},
  publisher={EDP Sciences}
}

@article{garcia2024galaxy,
  title={The Galaxy Activity, Torus, and Outflow Survey (GATOS)-III. Revealing the inner icy structure in local active galactic nuclei},
  author={Garc{\'\i}a-Bernete, I and Alonso-Herrero, A and Rigopoulou, D and Pereira-Santaella, Miguel and Shimizu, T and Davies, R and Donnan, FR and Roche, PF and Gonz{\'a}lez-Mart{\'\i}n, O and Almeida, C Ramos and others},
  journal={Astronomy \& Astrophysics},
  volume={681},
  pages={L7},
  year={2024},
  publisher={EDP Sciences}
}

@article{zhang2024galaxy,
  title={The Galaxy Activity, Torus, and Outflow Survey (GATOS). IV. Exploring Ionized Gas Outflows in Central Kiloparsec Regions of GATOS Seyferts},
  author={Zhang, Lulu and Packham, Chris and Hicks, Erin KS and Davies, Ric I and Shimizu, Taro T and Alonso-Herrero, Almudena and Mu{\~n}oz, Laura Hermosa and Garc{\'\i}a-Bernete, Ismael and Pereira-Santaella, Miguel and Audibert, Anelise and others},
  journal={The Astrophysical Journal},
  volume={974},
  number={2},
  pages={195},
  year={2024},
  publisher={IOP Publishing}
}

@article{garcia2022high,
  title={A high angular resolution view of the PAH emission in Seyfert galaxies using JWST/MRS data},
  author={Garc{\'\i}a-Bernete, I and Rigopoulou, D and Alonso-Herrero, A and Donnan, FR and Roche, PF and Pereira-Santaella, M and Labiano, A and de Arriba, L Peralta and Izumi, T and Almeida, C Ramos and others},
  journal={Astronomy \& Astrophysics},
  volume={666},
  pages={L5},
  year={2022},
  publisher={EDP Sciences}
}

@article{almeida2022diverse,
  title={The diverse cold molecular gas contents, morphologies, and kinematics of type-2 quasars as seen by ALMA},
  author={Ramos Almeida, C and Bischetti, M and Garc{\'\i}a-Burillo, S and Alonso-Herrero, A and Audibert, A and Cicone, Claudia and Feruglio, C and Tadhunter, CN and Pierce, JCS and Pereira-Santaella, M and others},
  journal={Astronomy \& Astrophysics},
  volume={658},
  pages={A155},
  year={2022},
  publisher={EDP Sciences}
}

@article{munoz2024biconical,
  title={A biconical ionised gas outflow and evidence of positive feedback in NGC 7172 uncovered by MIRI/JWST},
  author={Hermosa-Mu{\~n}oz, L  and Alonso-Herrero, A and Pereira-Santaella, M and Garc{\'\i}a-Bernete, I and Garc{\'\i}a-Burillo, S and Garc{\'\i}a-Lorenzo, B and Davies, R and Shimizu, T and Esparza-Arredondo, D and Hicks, EKS and others},
  journal={Astronomy \& Astrophysics},
  volume={690},
  pages={A350},
  year={2024},
  publisher={EDP Sciences}
}

@article{zhang2023interaction,
  title={The Interaction between AGN and Starburst Activity in the Circumnuclear Region of NGC 7469 as Viewed with JWST},
  author={Zhang, Lulu and Ho, Luis C},
  journal={The Astrophysical Journal Letters},
  volume={953},
  number={1},
  pages={L9},
  year={2023},
  publisher={IOP Publishing}
}

@article{pereira2022low,
  title={Low-power jet--interstellar medium interaction in NGC 7319 revealed by JWST/MIRI MRS},
  author={Pereira-Santaella, M and {\'A}lvarez-M{\'a}rquez, J and Garc{\'\i}a-Bernete, I and Labiano, A and Colina, L and Alonso-Herrero, A and Bellocchi, E and Garc{\'\i}a-Burillo, S and H{\"o}nig, SF and Almeida, C Ramos and others},
  journal={Astronomy \& Astrophysics},
  volume={665},
  pages={L11},
  year={2022},
  publisher={EDP Sciences}
}

@article{esparza2025molecular,
  title={Molecular gas stratification and disturbed kinematics in the Seyfert galaxy MCG-05-23-16 revealed by JWST and ALMA},
  author={Esparza-Arredondo, D and Almeida, C Ramos and Audibert, A and Pereira-Santaella, M and Garc{\'\i}a-Bernete, I and Garc{\'\i}a-Burillo, S and Shimizu, T and Davies, R and Mu{\~n}oz, L Hermosa and Alonso-Herrero, A and others},
  journal={Astronomy \& Astrophysics},
  volume={693},
  pages={A174},
  year={2025},
  publisher={EDP Sciences}
}

@article{harrison2024observational,
  title={Observational Tests of Active Galactic Nuclei Feedback: An Overview of Approaches and Interpretation},
  author={Harrison, Chris M and Ramos Almeida, Cristina},
  journal={Galaxies},
  volume={12},
  number={2},
  pages={17},
  year={2024},
  publisher={MDPI}
}

@article{garcia2021multiphase,
  title={Multiphase feedback processes in the Sy2 galaxy NGC 5643},
  author={Garc{\'\i}a-Bernete, Ismael and Alonso-Herrero, Almudena and Garc{\'\i}a-Burillo, Santiago and Pereira-Santaella, Miguel and Garc{\'\i}a-Lorenzo, B and Carrera, Francisco J and Rigopoulou, Dimitra and Almeida, C Ramos and Mart{\'\i}n, M Villar and Gonz{\'a}lez-Mart{\'\i}n, Omaira and others},
  journal={Astronomy \& Astrophysics},
  volume={645},
  pages={A21},
  year={2021},
  publisher={EDP Sciences}
}

@article{rieke2015mid,
  title={The mid-infrared instrument for the james webb space telescope, i: Introduction},
  author={Rieke, George H and Wright, GS and B{\"o}ker, T and Bouwman, Jeroen and Colina, Luis and Glasse, Alistair and Gordon, KD and Greene, TP and G{\"u}del, Manuel and Henning, Th and others},
  journal={Publications of the Astronomical Society of the Pacific},
  volume={127},
  number={953},
  pages={584},
  year={2015},
  publisher={IOP Publishing}
}

@article{eguchi2011suzaku,
  title={Suzaku View of the Swift/BAT Active Galactic Nuclei. III. Application of Numerical Torus Models to Two Nearly Compton Thick Active Galactic Nuclei (NGC 612 and NGC 3081)},
  author={Eguchi, Satoshi and Ueda, Yoshihiro and Awaki, Hisamitsu and Aird, James and Terashima, Yuichi and Mushotzky, Richard},
  journal={The Astrophysical Journal},
  volume={729},
  number={1},
  pages={31},
  year={2011},
  publisher={IOP Publishing}
}

@article{ueda2007suzaku,
  title={Suzaku observations of active galactic nuclei detected in the Swift BAT survey: Discovery of a “new type” of buried supermassive black holes},
  author={Ueda, Yoshihiro and Eguchi, Satoshi and Terashima, Yuichi and Mushotzky, Richard and Tueller, Jack and Markwardt, Craig and Gehrels, Neil and Hashimoto, Yasuhiro and Potter, Stephen},
  journal={The Astrophysical Journal},
  volume={664},
  number={2},
  pages={L79},
  year={2007},
  publisher={IOP Publishing}
}

@article{garcia2024deciphering,
  title={Deciphering the imprint of active galactic nucleus feedback in Seyfert galaxies-Nuclear-scale molecular gas deficits},
  author={Garc{\'\i}a-Burillo, Santiago and Hicks, EKS and Alonso-Herrero, A and Pereira-Santaella, M and Usero, A and Querejeta, M and Gonz{\'a}lez-Mart{\'\i}n, O and Delaney, D and Almeida, C Ramos and Combes, F and others},
  journal={Astronomy \& Astrophysics},
  volume={689},
  pages={A347},
  year={2024},
  publisher={EDP Sciences}
}

@article{gardner2023james,
  title={The James Webb space telescope mission},
  author={Gardner, Jonathan P and Mather, John C and Abbott, Randy and Abell, James S and Abernathy, Mark and Abney, Faith E and Abraham, John G and Abraham, Roberto and Abul-Huda, Yasin M and Acton, Scott and others},
  journal={Publications of the Astronomical Society of the Pacific},
  volume={135},
  number={1048},
  pages={068001},
  year={2023},
  publisher={IOP Publishing}
}

@article{rigby2023science,
  title={The science performance of JWST as characterized in commissioning},
  author={Rigby, Jane and Perrin, Marshall and McElwain, Michael and Kimble, Randy and Friedman, Scott and Lallo, Matt and Doyon, Ren{\'e} and Feinberg, Lee and Ferruit, Pierre and Glasse, Alistair and others},
  journal={Publications of the Astronomical Society of the Pacific},
  volume={135},
  number={1046},
  pages={048001},
  year={2023},
  publisher={IOP Publishing}
}

@article{garcia2016alma,
  title={ALMA resolves the torus of NGC 1068: continuum and molecular line emission},
  author={Garc{\'\i}a-Burillo, S and Combes, F and Almeida, C Ramos and Usero, A and Krips, M and Alonso-Herrero, A and Aalto, S and Casasola, VIVIANA and Hunt, Leslie Kipp and Mart{\'\i}n, S and others},
  journal={The Astrophysical Journal Letters},
  volume={823},
  number={1},
  pages={L12},
  year={2016},
  publisher={IOP Publishing}
}

@article{garcia2014molecular,
  title={Molecular line emission in ngc 1068 imaged with alma-i. an agn-driven outflow in the dense molecular gas},
  author={Garc{\'\i}a-Burillo, Santiago and Combes, Fran{\c{c}}oise and Usero, Antonio and Aalto, Susanne and Krips, Melanie and Viti, Serena and Alonso-Herrero, Almudena and Hunt, Leslie Kipp and Schinnerer, Eva and Baker, Andrew J and others},
  journal={Astronomy \& Astrophysics},
  volume={567},
  pages={A125},
  year={2014},
  publisher={EDP Sciences}
}

@article{garcia2019alma,
  title={ALMA images the many faces of the NGC 1068 torus and its surroundings},
  author={Garc{\'\i}a-Burillo, S and Combes, F and Almeida, C Ramos and Usero, A and Alonso-Herrero, A and Hunt, Leslie Kipp and Rouan, D and Aalto, Susanne and Querejeta, M and Viti, Serena and others},
  journal={Astronomy \& Astrophysics},
  volume={632},
  pages={A61},
  year={2019},
  publisher={EDP Sciences}
}

@article{gallimore2016high,
  title={High-velocity bipolar molecular emission from an AGN torus},
  author={Gallimore, Jack F and Elitzur, Moshe and Maiolino, Roberto and Marconi, Alessandro and O’Dea, Christopher P and Lutz, Dieter and Baum, Stefi A and Nikutta, Robert and Impellizzeri, CMV and Davies, Richard and others},
  journal={The Astrophysical Journal Letters},
  volume={829},
  number={1},
  pages={L7},
  year={2016},
  publisher={IOP Publishing}
}

@article{impellizzeri2019counter,
  title={Counter-rotation and High-velocity Outflow in the Parsec-scale Molecular Torus of NGC 1068},
  author={Impellizzeri, CM Violette and Gallimore, Jack F and Baum, Stefi A and Elitzur, Moshe and Davies, Richard and Lutz, Dieter and Maiolino, Roberto and Marconi, Alessandro and Nikutta, Robert and O’Dea, Christopher P and others},
  journal={The Astrophysical Journal Letters},
  volume={884},
  number={2},
  pages={L28},
  year={2019},
  publisher={IOP Publishing}
}

@article{imanishi2018alma,
  title={ALMA reveals an inhomogeneous compact rotating dense molecular torus at the NGC 1068 nucleus},
  author={Imanishi, Masatoshi and Nakanishi, Kouichiro and Izumi, Takuma and Wada, Keiichi},
  journal={The Astrophysical Journal Letters},
  volume={853},
  number={2},
  pages={L25},
  year={2018},
  publisher={IOP Publishing}
}

@article{harrison2018agn,
  title={AGN outflows and feedback twenty years on},
  author={Harrison, CM and Costa, T and Tadhunter, CN and Fl{\"u}tsch, A and Kakkad, D and Perna, M and Vietri, GIUSTINA},
  journal={Nature Astronomy},
  volume={2},
  number={3},
  pages={198--205},
  year={2018},
  publisher={Nature Publishing Group UK London}
}

@article{harrison2017impact,
  title={Impact of supermassive black hole growth on star formation},
  author={Harrison, CM},
  journal={Nature Astronomy},
  volume={1},
  number={7},
  pages={0165},
  year={2017},
  publisher={Nature Publishing Group UK London}
}

@article{de2023radio,
  title={A radio-jet-driven outflow in the Seyfert 2 galaxy NGC 2110?},
  author={de Arriba, L Peralta and Alonso-Herrero, A and Garc{\'\i}a-Burillo, S and Garc{\'\i}a-Bernete, I and Villar-Mart{\'\i}n, M and Garc{\'\i}a-Lorenzo, B and Davies, R and Rosario, DJ and H{\"o}nig, SF and Levenson, NA and others},
  journal={Astronomy \& Astrophysics},
  volume={675},
  pages={A58},
  year={2023},
  publisher={EDP Sciences}
}

@article{holden2024alma,
  title={ALMA reveals a compact and massive molecular outflow driven by the young AGN in a nearby ULIRG},
  author={Holden, Luke R and Tadhunter, Clive and Audibert, Anelise and Oosterloo, Tom and Ramos Almeida, Cristina and Morganti, Raffaella and Pereira-Santaella, Miguel and Lamperti, Isabella},
  journal={Monthly Notices of the Royal Astronomical Society},
  volume={530},
  number={1},
  pages={446--456},
  year={2024},
  publisher={Oxford University Press}
}

@article{riffel2025blowing,
  title={Blowing Star Formation Away in AGN Hosts (BAH). II. Investigating the Origin of the H2 Emission Excess in Nearby Galaxies with JWST MIRI},
  author={Riffel, Rogemar A and Souza-Oliveira, Gabriel L and Costa-Souza, Jos{\'e} Henrique and Zakamska, Nadia L and Storchi-Bergmann, Thaisa and Riffel, Rog{\'e}rio and Bianchin, Marina},
  journal={The Astrophysical Journal},
  volume={982},
  number={2},
  pages={69},
  year={2025},
  publisher={IOP Publishing}
}

@article{schmitt2001jet,
  title={Jet Directions in Seyfert Galaxies: Radio Continuum ImagingData},
  author={Schmitt, HR and Ulvestad, JS and Antonucci, RRJ and Kinney, AL},
  journal={The Astrophysical Journal Supplement Series},
  volume={132},
  number={2},
  pages={199},
  year={2001},
  publisher={IOP Publishing}
}

@misc{NVAS,
  author = "{NRAO/VLA Archive Survey}",
  title = "{NRAO/VLA Archive Survey Image}",
  year = {2007},
  note = "Image credit: NRAO/VLA Archive Survey, (c) 2005-2007 AUI/NRAO",
  url = {https://archive.nrao.edu/nvas/}
}

@article{rosario2018llama,
  title={LLAMA: normal star formation efficiencies of molecular gas in the centres of luminous Seyfert galaxies},
  author={Rosario, DJ and Burtscher, Leonard and Davies, Richard I and Koss, Michael and Ricci, Claudio and Lutz, Dieter and Riffel, R and Alexander, David M and Genzel, Reinhard and Hicks, EH and others},
  journal={Monthly Notices of the Royal Astronomical Society},
  volume={473},
  number={4},
  pages={5658--5679},
  year={2018},
  publisher={Oxford University Press}
}

@article{veron2006catalogue,
  title={A catalogue of quasars and active nuclei},
  author={V{\'e}ron-Cetty, M-P and V{\'e}ron, Philippe},
  journal={Astronomy \& Astrophysics},
  volume={455},
  number={2},
  pages={773--777},
  year={2006},
  publisher={EDP Sciences}
}

@article{bessiere2022spatially,
  title={Spatially resolved evidence of the impact of quasar-driven outflows on recent star formation: the case of Mrk 34},
  author={Bessiere, Patricia S and Ramos Almeida, Cristina},
  journal={Monthly Notices of the Royal Astronomical Society: Letters},
  volume={512},
  number={1},
  pages={L54--L59},
  year={2022},
  publisher={Oxford University Press}
}

@article{SilkRees1998,
  author  = {Silk, Joseph and Rees, Martin J.},
  title   = {Quasars and galaxy formation},
  journal = {Astronomy and Astrophysics},
  volume  = {331},
  pages   = {L1--L4},
  year    = {1998},
  url     = {https://articles.adsabs.harvard.edu/pdf/1998A%26A...331L...1S}
}

@article{fischer2016gemini,
  title={Gemini Near Infrared Field Spectrograph observations of the Seyfert 2 galaxy Mrk 573: in situ acceleration of ionized and molecular gas off fueling flows},
  author={Fischer, Travis C and Machuca, Camilo and Diniz, Marlon Rodrigo and Crenshaw, Daniel Michael and Kraemer, Steven B and Riffel, Rogemar Andr{\'e} and Schmitt, Henrique Roberto and Baron, Fabien and Storchi-Bergmann, Thaisa and Straughn, Amber N and others},
  journal={The Astrophysical Journal},
  volume={834},
  number={1},
  pages={30},
  year={2016},
  publisher={IOP Publishing}
}

@article{fischer2018hubble,
  title={Hubble Space Telescope observations of extended [O III] $\lambda$ 5007 emission in nearby QSO2s: new constraints on AGN host galaxy interaction},
  author={Fischer, Travis C and Kraemer, SB and Schmitt, Henrique Roberto and Micchi, LF Longo and Crenshaw, Daniel Michael and Revalski, Mitchell and Vestergaard, Marianne and Elvis, Martin and Gaskell, CM and Hamann, Fred and others},
  journal={The Astrophysical Journal},
  volume={856},
  number={2},
  pages={102},
  year={2018},
  publisher={IOP Publishing}
}

@article{khalatyan2008agn,
  title={Is AGN feedback necessary to form red elliptical galaxies?},
  author={Khalatyan, A and Cattaneo, Andrea and Schramm, M and Gottl{\"o}ber, S and Steinmetz, M and Wisotzki, L},
  journal={Monthly Notices of the Royal Astronomical Society},
  volume={387},
  number={1},
  pages={13--30},
  year={2008},
  publisher={Blackwell Publishing Ltd Oxford, UK}
}

@article{hopkins2005black,
  title={Black holes in galaxy mergers: Evolution of quasars},
  author={Hopkins, Philip F and Hernquist, Lars and Cox, Thomas J and Di Matteo, Tiziana and Martini, Paul and Robertson, Brant and Springel, Volker},
  journal={The Astrophysical Journal},
  volume={630},
  number={2},
  pages={705},
  year={2005},
  publisher={IOP Publishing}
}

@article{crenshaw2010geometry,
  title={The Geometry of Mass Outflows and Fueling Flows in the Seyfert 2 Galaxy MRK 3},
  author={Crenshaw, DM and Kraemer, SB and Schmitt, HR and Jaff{\'e}, YL and Deo, RP and Collins, NR and Fischer, TC},
  journal={The Astronomical Journal},
  volume={139},
  number={3},
  pages={871},
  year={2010},
  publisher={IOP Publishing}
}

@article{crenshaw2012feedback,
  title={Feedback from mass outflows in nearby active galactic nuclei. I. Ultraviolet and X-ray absorbers},
  author={Crenshaw, D Michael and Kraemer, Steven B},
  journal={The Astrophysical Journal},
  volume={753},
  number={1},
  pages={75},
  year={2012},
  publisher={IOP Publishing}
}

@article{fischer2011hubble,
  title={Hubble Space Telescope Observations of the Double-peaked Emission Lines in the Seyfert Galaxy Markarian 78: Mass Outflows from a Single Active Galactic Nucleus},
  author={Fischer, TC and Crenshaw, DM and Kraemer, SB and Schmitt, HR and Mushotsky, RF and Dunn, JP},
  journal={The Astrophysical Journal},
  volume={727},
  number={2},
  pages={71},
  year={2011},
  publisher={IOP Publishing}
}

@article{jones1994grain,
  title={Grain destruction in shocks in the interstellar medium},
  author={Jones, AP and Tielens, AGGM and Hollenbach, DJ and McKee, CF},
  journal={Astrophysical Journal, Part 1 (ISSN 0004-637X), vol. 433, no. 2, p. 797-810},
  volume={433},
  pages={797--810},
  year={1994}
}

@article{jones1996grain,
  title={Grain shattering in shocks: The interstellar grain size distribution},
  author={Jones, AP and Tielens, AGGM and Hollenbach, DJ},
  journal={Astrophysical Journal v. 469, p. 740},
  volume={469},
  pages={740},
  year={1996}
}

@article{tielens1994physics,
  title={The physics of grain-grain collisions and gas-grain sputtering in interstellar shocks},
  author={Tielens, AG and McKee, CF and Seab, CG and Hollenbach, DJ},
  journal={Astrophys J},
  pages={321},
  year={1994}
}

@article{muller2011outflows,
  title={Outflows from active galactic nuclei: kinematics of the narrow-line and coronal-line regions in Seyfert galaxies},
  author={M{\"u}ller-S{\'a}nchez, F and Prieto, MA and Hicks, EKS and Vives-Arias, H and Davies, RI and Malkan, M and Tacconi, LJ and Genzel, R},
  journal={The Astrophysical Journal},
  volume={739},
  number={2},
  pages={69},
  year={2011},
  publisher={IOP Publishing}
}

@article{alonso2025miconic,
  title={MICONIC: JWST/MIRI MRS reveals a fast ionized gas outflow in the central region of Centaurus A},
  author={Alonso-Herrero, A and Mu{\~n}oz, L and Labiano, A and Guillard, P and Garc{\'\i}a-Mar{\'\i}n, M and Dicken, D and Garc{\'\i}a-Burillo, S and Pantoni, L and Buiten, V and Colina, L and others},
  journal={arXiv preprint arXiv:2506.15286},
  year={2025}
}

@article{zubovas2020intermittent,
  title={Intermittent AGN episodes drive outflows with a large spread of observable loading factors},
  author={Zubovas, Kastytis and Nardini, Emanuele},
  journal={Monthly Notices of the Royal Astronomical Society},
  volume={498},
  number={3},
  pages={3633--3647},
  year={2020},
  publisher={Oxford University Press}
}

@inproceedings{bergmann2012resolved,
  title={Resolved Outflows in Nearby AGN from Integral Field Spectroscopy},
  author={Bergmann, Thaisa Storchi},
  booktitle={AGN Winds in Charleston},
  volume={460},
  pages={133},
  year={2012}
}

@article{shimizu2019multiphase,
  title={The multiphase gas structure and kinematics in the circumnuclear region of NGC 5728},
  author={Shimizu, T Taro and Davies, RI and Lutz, D and Burtscher, L and Lin, M and Baron, D and Davies, RL and Genzel, R and Hicks, EKS and Koss, M and others},
  journal={Monthly Notices of the Royal Astronomical Society},
  volume={490},
  number={4},
  pages={5860--5887},
  year={2019},
  publisher={Oxford University Press}
}

@article{smajic2012unveiling,
  title={Unveiling the nucleus of NGC 7172},
  author={Smaji{\'c}, Semir and Fischer, Sebastian and Zuther, Jens and Eckart, A},
  journal={Astronomy \& Astrophysics},
  volume={544},
  pages={A105},
  year={2012},
  publisher={EDP Sciences}
}

@article{burtscher2021llama,
  title={LLAMA: Stellar populations in the nuclei of ultra-hard X-ray-selected AGN and matched inactive galaxies},
  author={Burtscher, Leonard and Davies, Richard I and Shimizu, TT and Riffel, Rog{\'e}rio and Rosario, DJ and Hicks, Erin KS and Lin, M-Y and Riffel, Rogemar Andr{\'e} and Schartmann, Marc and Schnorr-M{\"u}ller, A and others},
  journal={Astronomy \& Astrophysics},
  volume={654},
  pages={A132},
  year={2021},
  publisher={EDP Sciences}
}

@article{lien2025157,
  title={The 157-month Swift-BAT All-Sky Hard X-Ray Survey},
  author={Lien, Amy Y and Krimm, Hans and Markwardt, Craig and Oh, Kyuseok and Marcotulli, Lea and Mushotzky, Richard and Collins, Nicholas R and Barthelmy, Scott and Baumgartner, Wayne H and Cenko, S Bradley and others},
  journal={arXiv preprint arXiv:2506.04109},
  year={2025}
}

@article{ma2020extended,
  title={Is extended hard X-ray emission ubiquitous in Compton-thick AGN?},
  author={Ma, Jingzhe and Elvis, Martin and Fabbiano, G and Balokovi{\'c}, Mislav and Maksym, W Peter and Jones, Mackenzie L and Risaliti, Guido},
  journal={The Astrophysical Journal},
  volume={900},
  number={2},
  pages={164},
  year={2020},
  publisher={IOP Publishing}
}

@article{davies2015insights,
  title={Insights on the dusty torus and neutral torus from optical and X-ray obscuration in a complete volume limited hard X-ray AGN sample},
  author={Davies, Richard I and Burtscher, Leonard and Rosario, David and Storchi-Bergmann, Thaisa and Contursi, Alessandra and Genzel, Reinhard and Gracia-Carpio, Javier and Hicks, E and Janssen, Annemieke and Koss, Michael and others},
  journal={The Astrophysical Journal},
  volume={806},
  number={1},
  pages={127},
  year={2015},
  publisher={IOP Publishing}
}

@article{audibert2023jet,
  title={Jet-induced molecular gas excitation and turbulence in the Teacup},
  author={Audibert, A and Almeida, C Ramos and Garc{\'\i}a-Burillo, S and Combes, Francoise and Bischetti, M and Meenakshi, M and Mukherjee, D and Bicknell, G and Wagner, AY},
  journal={Astronomy \& Astrophysics},
  volume={671},
  pages={L12},
  year={2023},
  publisher={EDP Sciences}
}

@article{ruschel2021agnifs,
  title={AGNIFS survey of local AGN: GMOS-IFU data and outflows in 30 sources},
  author={Ruschel-Dutra, D and Storchi-Bergmann, T and Schnorr-M{\"u}ller, A and Riffel, RA and Dall’Agnol de Oliveira, B and Lena, D and Robinson, A and Nagar, N and Elvis, M},
  journal={Monthly Notices of the Royal Astronomical Society},
  volume={507},
  number={1},
  pages={74--89},
  year={2021},
  publisher={Oxford University Press}
}

@article{herrero2024miconic,
  title={MICONIC: JWST/MIRI MRS observations of the nuclear and circumnuclear regions of Mrk 231},
  author={Alonso-Herrero, A and Mu{\~n}oz, L Hermosa and Labiano, Alvaro and Guillard, Pierre and Buiten, Victorine A and Dicken, Daniel and Van Der Werf, P and {\'A}lvarez-M{\'a}rquez, Javier and B{\"o}ker, Torsten and Colina, Luis and others},
  journal={Astronomy \& Astrophysics},
  volume={690},
  pages={A95},
  year={2024},
  publisher={EDP Sciences}
}

@misc{imanishi2025almasubparsecresolutiondense,
      title={ALMA Sub-parsec Resolution Dense Molecular Line Observations of the NGC 1068 Nucleus}, 
      author={Masatoshi Imanishi and Bernd Vollmer and Yoshiaki Hagiwara and Kouichiro Nakanishi and Takuma Izumi and Nozomu Kawakatu},
      year={2025},
      eprint={2511.04772},
      archivePrefix={arXiv},
      primaryClass={astro-ph.GA},
      url={https://arxiv.org/abs/2511.04772}, 
}

@article{riffel2025impact,
  title={Impact of AGN and nuclear star formation on the ISM turbulence of galaxies: Insights from JWST/MIRI spectroscopy},
  author={Riffel, Rogemar A and Colina, Luis and Costa-Souza, Jos{\'e} Henrique and Mainieri, Vincenzo and Santaella, Miguel Pereira and Dors, Oli L and Garc{\'\i}a-Bernete, Ismael and Alonso-Herrero, Almudena and Audibert, Anelise and Bellocchi, Enrica and others},
  journal={arXiv preprint arXiv:2510.02517},
  year={2025}
}

@ARTICLE{Jenkins2009ApJ,
       author = {{Jenkins}, Edward B.},
        title = "{A Unified Representation of Gas-Phase Element Depletions in the Interstellar Medium}",
      journal = {\apj},
         year = 2009,
        month = aug,
       volume = {700},
       number = {2},
        pages = {1299-1348},
          doi = {10.1088/0004-637X/700/2/1299},
archivePrefix = {arXiv},
       eprint = {0905.3173},
 primaryClass = {astro-ph.GA},
       adsurl = {https://ui.adsabs.harvard.edu/abs/2009ApJ...700.1299J}
}

@ARTICLE{Sofia1998,
       author = {{Sofia}, Ulysses J. and {Jenkins}, Edward B.},
        title = "{Interstellar Medium Absorption Profile Spectrograph Observations of Interstellar Neutral Argon and the Implications for Partially Ionized Gas}",
      journal = {\apj},
         year = 1998,
        month = may,
       volume = {499},
       number = {2},
        pages = {951-965},
          doi = {10.1086/305684},
archivePrefix = {arXiv},
       eprint = {astro-ph/9712260},
 primaryClass = {astro-ph},
       adsurl = {https://ui.adsabs.harvard.edu/abs/1998ApJ...499..951S}
}

@article{verma2003mid,
  title={A mid-infrared spectroscopic survey of starburst galaxies: Excitation and abundances},
  author={Verma, Aprajita and Lutz, Dieter and Sturm, Eckhard and Sternberg, Amiel and Genzel, Reinhard and Vacca, William},
  journal={Astronomy \& Astrophysics},
  volume={403},
  number={3},
  pages={829--846},
  year={2003},
  publisher={EDP Sciences}
}

@article{amayo2021ionization,
  title={Ionization correction factors and dust depletion patterns in giant H II regions},
  author={Amayo, A and Delgado-Inglada, G and Stasi{\'n}ska, G},
  journal={Monthly Notices of the Royal Astronomical Society},
  volume={505},
  number={2},
  pages={2361--2376},
  year={2021},
  publisher={Oxford University Press}
}

@article{lodders2003solar,
  title={Solar system abundances and condensation temperatures of the elements},
  author={Lodders, Katharina},
  journal={The Astrophysical Journal},
  volume={591},
  number={2},
  pages={1220},
  year={2003},
  publisher={IOP Publishing}
}

@article{antonucci1993unified,
  title={Unified models for active galactic nuclei and quasars},
  author={Antonucci, Robert},
  journal={In: Annual review of astronomy and astrophysics. Vol. 31 (A94-12726 02-90), p. 473-521.},
  volume={31},
  pages={473--521},
  year={1993}
}

@article{urry1995unified,
  title={Unified schemes for radio-loud active galactic nuclei},
  author={Urry, C Megan and Padovani, Paolo},
  journal={Publications of the Astronomical Society of the Pacific},
  volume={107},
  number={715},
  pages={803},
  year={1995},
  publisher={IOP Publishing}
}

@article{delaney2025excitation,
  title={Excitation of Molecular Hydrogen in Seyferts: NGC 5506 and NGC 3081},
  author={Delaney, Daniel E and Hicks, Erin KS and Zhang, Lulu and Packham, Chris and Davies, Ric and Santaella, Miguel Pereira and Bellocchi, Enrica and Levenson, Nancy A and Campbell, Steph and Rosario, David J and others},
  journal={The Astrophysical Journal},
  volume={993},
  number={2},
  pages={217},
  year={2025},
  publisher={IOP Publishing}
}

@article{ogle2007shocked,
  title={Shocked molecular hydrogen in the 3C 326 Radio Galaxy system},
  author={Ogle, Patrick and Antonucci, Robert and Appleton, PN and Whysong, David},
  journal={The Astrophysical Journal},
  volume={668},
  number={2},
  pages={699},
  year={2007},
  publisher={IOP Publishing}
}

@article{fischer2019dissection,
  title={A Dissection of Spatially Resolved AGN Feedback across the Electromagnetic Spectrum},
  author={Fischer, Travis and Smith, Krista Lynne and Kraemer, Steve and Schmitt, Henrique and Crenshaw, D Michael and Koss, Michael and Mushotzky, Richard and Larson, Kirsten and Rigby, Jane and others},
  journal={The Astrophysical Journal},
  volume={887},
  number={2},
  pages={200},
  year={2019},
  publisher={IOP Publishing}
}

@article{fischer2023no,
  title={No small-scale radio jets here: multiepoch observations of radio continuum structures in NGC 1068 with the VLBA},
  author={Fischer, Travis C and Johnson, Megan C and Secrest, Nathan J and Crenshaw, D Michael and Kraemer, Steven B},
  journal={The Astrophysical Journal},
  volume={953},
  number={1},
  pages={87},
  year={2023},
  publisher={IOP Publishing}
}

@article{zakamska2014quasar,
  title={Quasar feedback and the origin of radio emission in radio-quiet quasars},
  author={Zakamska, Nadia L and Greene, Jenny E},
  journal={Monthly Notices of the Royal Astronomical Society},
  volume={442},
  number={1},
  pages={784--804},
  year={2014},
  publisher={Oxford University Press}
}

@article{orienti2010radio,
  title={Radio structures of the nuclei of nearby Seyfert galaxies and the nature of the missing diffuse emission},
  author={Orienti, Monica and Prieto, MA},
  journal={Monthly Notices of the Royal Astronomical Society},
  volume={401},
  number={4},
  pages={2599--2610},
  year={2010},
  publisher={Blackwell Publishing Ltd Oxford, UK}
}

@article{maiolino2017star,
  title={Star formation inside a galactic outflow},
  author={Maiolino, Roberto and Russell, HR and Fabian, Andrew C and Carniani, Stefano and Gallagher, R and Cazzoli, S and Arribas, S and Belfiore, F and Bellocchi, E and Colina, Luis and others},
  journal={Nature},
  volume={544},
  number={7649},
  pages={202--206},
  year={2017},
  publisher={Nature Publishing Group UK London}
}

@article{birzan2008radiative,
  title={Radiative efficiency and content of extragalactic radio sources: toward a universal scaling relation between jet power and radio power},
  author={B{\^\i}rzan, L and McNamara, BR and Nulsen, PEJ and Carilli, CL and Wise, MW},
  journal={The Astrophysical Journal},
  volume={686},
  number={2},
  pages={859},
  year={2008},
  publisher={IOP Publishing}
}

@article{zhang2025theoretical,
  title={Theoretical Diagnostics for the Physical Conditions in Active Galactic Nuclei under the View of JWST},
  author={Zhang, Lulu and Davies, Ric I and Packham, Chris and Hicks, Erin KS and Delaney, Daniel E and Pereira-Santaella, Miguel and Mu{\~n}oz, Laura Hermosa and Garc{\'\i}a-Bernete, Ismael and Ricci, Claudio and Rigopoulou, Dimitra and others},
  journal={The Astrophysical Journal Supplement Series},
  volume={280},
  number={2},
  pages={65},
  year={2025},
  publisher={IOP Publishing}
}

@article{hunt2025interstellar,
  title={The Interstellar Medium in I Zw 18 Seen with JWST/MIRI. II. Warm Molecular Hydrogen and Warm Dust},
  author={Hunt, LK and Draine, BT and Navarro, MG and Aloisi, A and Vaught, RJ Rickards and Adamo, A and Annibali, F and Calzetti, D and Hernandez, S and James, BL and others},
  journal={The Astrophysical Journal},
  volume={993},
  number={1},
  pages={84},
  year={2025},
  publisher={IOP Publishing}
}

@article{venturi2021magnum,
  title={MAGNUM survey: Compact jets causing large turmoil in galaxies-Enhanced line widths perpendicular to radio jets as tracers of jet-ISM interaction},
  author={Venturi, Giacomo and Cresci, Giovanni and Marconi, Alessandro and Mingozzi, M and Nardini, E and Carniani, Stefano and Mannucci, Filippo and Marasco, A and Maiolino, Roberto and Perna, Marianna and others},
  journal={Astronomy \& Astrophysics},
  volume={648},
  pages={A17},
  year={2021},
  publisher={EDP Sciences}
}

@article{krol2026probing,
  title={Probing Active Galactic Nuclei--Interstellar Medium Feedback through Extended X-Ray Emission in ESO 137-G034},
  author={Kr{\'o}l, D{\L} and Fabbiano, G and Elvis, M and Trindade Falc{\~a}o, A and Middei, R and Rosario, D and Davies, R and Shimizu, T and Hill, D},
  journal={The Astrophysical Journal},
  volume={998},
  number={1},
  pages={135},
  year={2026},
  publisher={The American Astronomical Society}
}

@article{haidar2026gatos,
  title={GATOS XI: Excess dust heating in the Narrow Line Regions of nearby AGN revealed with JWST/MIRI},
  author={Haidar, Houda and Rosario, David J and Garc{\'\i}a-Bernete, Ismael and Alonso-Herrero, Almudena and Audibert, Anelise and Campbell, Steph and Harrison, Chris M and Costa, Tiago and Mu{\~n}oz, Laura Hermosa and Combes, Fran{\c{c}}oise and others},
  journal={Monthly Notices of the Royal Astronomical Society},
  pages={stag069},
  year={2026},
  publisher={Oxford University Press}
}
\bibliographystyle{aasjournal}



\end{document}